\documentclass[12pt,preprint]{aastex}

\usepackage{natbib}
\usepackage{pdflscape}

\usepackage{afterpage}
\usepackage{capt-of}

\def\gcc{\hbox{\rm\hskip.35em  g cm}$^{-3}$}

\def\ergcc{\hbox{\rm\hskip.35em  erg cm}$^{-3}$}

\def\eg{{\it e.g.}}
\def\lap{\hbox{${_{\displaystyle<}\atop^{\displaystyle\sim}}$}}
\def\gap{\hbox{${_{\displaystyle>}\atop^{\displaystyle\sim}}$}}

\newcommand{\ev}[1]{\ensuremath{\langle #1\rangle}}

\newcommand{\rbf}       {\mbox{\boldmath$r$}}
\newcommand{\ubf}       {\mbox{\boldmath$u$}}

\newcommand{\Bbf}       {\mbox{\boldmath$B$}}

\def\eg{{\it e.g.}}
\def\lap{\hbox{${_{\displaystyle<}\atop^{\displaystyle\sim}}$}}
\def\gap{\hbox{${_{\displaystyle>}\atop^{\displaystyle\sim}}$}}

\usepackage{graphicx}
\usepackage{subfigure}
\begin{document}

\title{TORSIONAL OSCILLATIONS OF A MAGNETAR WITH A TANGLED MAGNETIC 
FIELD\footnote{Supplementary materials to this Letter appear 
in \citet{lv16s}.}}

\author{Bennett Link}
\author{C. Anthony van Eysden}
\affil{Department of Physics, Montana State University, Bozeman,  MT
59717, USA:}
\email{link@montana.edu}
\email{anthonyvaneysden@montana.edu}

\date{\bigskip\centerline{\sl Astrophysical Journal Letters, in press}}

\begin{abstract}
Motivated by stability considerations and observational evidence, we
argue that magnetars possess highly-tangled internal magnetic fields.
We propose that the quasi-periodic oscillations (QPOs) seen to
accompany giant flares can be explained as torsional modes supported
by a tangled magnetic field, and we present a simple model that
supports this hypothesis for SGR 1900+14. Taking the strength of the
tangle as a free parameter, we find that the magnetic energy in the
tangle must dominate that in the dipolar component by a factor of
$\sim 14$ to accommodate the observed 28 Hz QPO. Our simple model
provides useful scaling relations for how the QPO spectrum depends on
the bulk properties of the neutron star and the tangle strength.  The
energy density in the tangled field inferred for SGR 1900+14 renders
the crust nearly dynamically irrelevant, a significant simplification
for study of the QPO problem.  The predicted spectrum is about three
times denser than observed, which could be explained by preferential
mode excitation or beamed emission. We emphasize that field tangling
is needed to stabilize the magnetic field, so should not be ignored in
treatment of the QPO problem.

\end{abstract}


\section{Introduction}

Soft-gamma repeaters (SGRs) are strongly-magnetized neutron stars that
produce frequent, short-duration bursts ($\lap 1$ s) of $\lap 10^{41}$
ergs in hard x-ray and soft gamma-rays. SGRs occasionally produce
giant flares that last $\sim 100$ s; the first giant flare to be
detected occurred in SGR 0526-66 on 5 March, 1979
\citep{barat_etal79,mazets_etal79,cline_etal80}, releasing 
$\sim 2\times 10^{45}$ erg \citep{fenimore_etal96}. The August
27th 1998 giant flare from SGR 1900+14 liberated $\gap 4\times
10^{43}$ erg, with a rise time of $<4$ ms
\citep{hurley_etal99,feroci_etal99}. The duration of the initial peak
was $\sim 1$ s \citep{hurley_etal99}. On December 27, 2004, SGR
1806-20 produced the largest flare yet recorded, with a total energy
yield of $\gap 4\times 10^{46}$ ergs.\footnote{These energy estimates
assume isotropic emission.} In both short bursts and in giant flares,
the peak luminosity is reached in under 10 ms.  Measured spin down
parameters imply surface dipole fields of $6\times 10^{14}$ G for SGR
0526-66 \citep{tiengo_etal09}, $7\times 10^{14}$ G for SGR 1900+14
\citep{mereghetti_etal06}, and $2\times 10^{15}$ G for SGR 1806-20
\citep{nakagawa_etal08}, establishing these objects as magnetars.

The giant flares in SGR 1806-20 (hereafter SGR 1806) and SGR 1900+14
(hereafter SGR 1900) showed rotationally phase-dependent,
quasi-periodic oscillations (QPOs). QPOs in SGR 1806 were detected
at 18$\pm 2$ Hz, 26$\pm 3$ Hz, 30$\pm 4$ Hz, 93$\pm 2$ Hz, 150$\pm
17$ Hz, 626$\pm 2$ Hz, and 1837$\pm 5$ Hz
\citep{israel_etal05,ws06,sw06,hambaryan_etal11}. 
QPOs in the giant flare of SGR 1900 were detected at 28$\pm 2$ Hz, 
53 $\pm 5$ Hz, 84 Hz (width unmeasured), and 155$\pm 6$ Hz
\citep{sw05}. 
Recently, oscillations at 57$\pm 5$ Hz were identified in the
short bursts of SGR 1806 \citep{huppenkothen_etal14a}, and
at 93$\pm 12$ Hz, 127$\pm 10$ Hz, and possibly 260 Hz in SGR J1550-5418
\citep{huppenkothen_etal14b}.\footnote{\citet{elib10} reported
evidence for oscillations in the
short, recurring bursts of SGR 1806, but this analysis was shown by
\citet{huppenkothen_etal13} to be flawed.}
To summarize, SGRs 1806 and 1900 have QPOs
that begin at about 20 Hz, with a spacing of some tens of
Hz below 160 Hz, and that are sharp with typical widths of 2-4
Hz. 

The observed QPOs are generally attributed to oscillations of the star
excited by an explosion of magnetic origin that creates the flare. The
oscillating stellar surface should modulate the charge density in the
magnetosphere, creating variations in the optical depth for resonant
Compton scattering of the hard x-rays that accompany the flare
\citep{timokhin_etal08,dw12}.  In this connection, the problem of
finding the oscillatory modes for a strongly-magnetized neutron star
has received much attention, and has proven to be a formidable
problem. To make the problem tractable, most theoretical treatments of
the QPO problem have assumed smooth field geometries, usually dipolar
or variants
(\eg, \citealt{levin06,gsa06,levin07,sotani_etal08a,sotani_etal08b,cerda_etal09,cbk09,csf09,ck11,vl11,gabler_etal11,ck11,gabler_etal12,vl12,pl13,gabler_etal13,gabler_etal_sf13,gabler_etal14}).
Smooth field geometries support a problematic Alfv\'en continuum that
couples to the discrete natural spectrum of the crust. As
pointed out by \citet{levin06}, if energy is deposited in the crust at
one of the natural frequencies of the crust, and this frequency lies
within a portion of the core continuum, the energy is lost to the core
continuum in less than 0.1 s as the entire core continuum is
excited. The crust excitation is effectively damped through {\em
resonant absorption}, a familiar process in MHD; see
\eg, \citet{gp04}. The problem has been addressed by assuming field
geometries with gaps in the Alfv\'en continuum. Under this assumption,
long-lived quasi-normal modes can exist inside the gaps or near the
edges of the Alfv\'en continuum. \cite{vl11} showed for a ``box''
neutron star that introduction of a magnetic tangle breaks the
Alfv\'en continuum. \cite{lv15} showed that for magnetic tangling in a
spherical neutron star the problematic Alfv\'en continuum disappears.\footnote{\cite{sotani15}
added general relativity in the treatment of the 
magnetic field, and
confirmed some of the results of \cite{lv15}.}
They found that the star acquires discrete normal modes, and
quantified the mode spacing. It is clear from these investigations
that the unknown magnetic field geometry is the most important
ingredient in determining the oscillation spectrum of a magnetar.

As no model presented so far has provided good quantitative agreement
with observed QPOs, we take a new direction in this Letter. We begin
by arguing that stability considerations and observational evidence
show that magnetars do not possess the smooth fields considered in
most previous work, but rather have highly tangled fields. We propose
that magnetar QPOs represent torsional normal modes that are supported
by the magnetic tangle, and we present a simple model that supports
this hypothesis. Keeping the energy in the magnetic tangle as a free
parameter, we adjust this parameter to accommodate the 28 Hz QPO
observed in SGR 1900 while maintaining consistency with QPOs observed
at higher frequencies. We obtain a rough {\em measurement} of the
energy density in the tangled field to be $\sim 14$ times that in the
dipole field. Our model, though simple, is the first to give
reasonable quantitative agreement with the data. Our model also
provides useful scaling relations for the frequencies of the QPOs on
bulk neutron star parameters and provides insight into the problem
that might not emerge so clearly from more detailed numerical
simulations. In particular, the model shows that if strong field
tangling occurs, the normal-mode spectrum of a magnetar is determined
principally by field tangling, and less so by crust rigidity, the
dipole field, relativistic effects, and detailed stellar structure.
We conclude that the effects of a tangled field cannot be neglected in
the QPO problem, and we outline what we see to be interesting research
directions on this issue. 

\section{Theoretical and observational evidence for field tangling}

A pure dipole field is unstable, and a strong toroidal field is needed
to stabilize the field
\citep{fr77,bs06}. Purely toroidal fields are also unstable
\citep{wright73,tayler73}. 
There has been
considerable progress recently on the identification of magnetic
equilibria. \citet{bn06} found a ``twisted torus'' configuration,
which consists of torus of flux near the magnetic equator that
stabilizes the linked poloidal plus toroidal configuration. The
twisted torus is topologically distinct from any poloidal field, or
twisted poloidal field, in the sense that the twisted torus cannot be
continuously deformed into a dipole field - the field is {\em
tangled}. This topological complexity is required to
establish hydromagnetic stability. 
Simulations by \citet{braithwaite08} show
that the evolution of the magnetic field from initially-turbulent
configurations can evolve to configurations other than the 
twisted torus, generally non-axisymmetric equilibria
with highly tangled fields; see, \eg, Fig. 12 of that
paper. \citet{braithwaite09} studied the relative strengths of the
poloidal and toroidal components in stable, axi-symmetric
configurations, and found that the energy in the toroidal 
component typically exceeds that in the poloidal component. 
By what factor the toroidal energy exceeds the poloidal energy 
in an actual neutron star
depends on initial conditions and the equation of state; 
\cite{braithwaite09} finds examples in which this ratio is 10-20, and he
argues that this ratio could plausibly be $\sim 10^3$ since a
proto-magnetar should be in a highly turbulent state that winds up the
natal field \citep{td93,brandenburg_etal05}. In this process, energy
injected at some scale propagates down to the dissipative scale as
well as up to large scales, giving a large-scale mean field with
complicated structure at many scales.

The chief conclusion of these theoretical studies is that a
topologically distinct tangle is needed to stabilize the dipolar
component. Most theoretical work on QPOs has assumed simple field geometries
that are demonstrably unstable.  

Observational evidence that the internal fields of neutron stars
are highly tangled can be found in the `low-field SGRs'. In these
objects, the interior fields must be stronger than the inferred dipole
fields in order to power observed burst activity. SGR 0418+5729 has a
dipole field inferred from spin-down of $\sim 6\times 10^{12}$ G,
\citep{esposito_etal10,ret10,horst_etal10,turolla_etal11}. Two other
examples are Swift J1822-1606, with an inferred dipole field of $\sim
3\times 10^{13}$ G \citep{rea_etal12}, and 3XMM J1852+0033
\citep{rvi14}, with an inferred dipole field of less than $4\times
10^{13}$ G.  Energetics indicate that the interior field consists of
strong multipolar components, while stability considerations require
these components to be tangled \citep{braithwaite08,braithwaite09}.

\section{A simple model of QPOs}

\label{model}

Much of the work cited above on the QPO problem has included realistic
stellar structure, specific magnetic field geometries, and the effects
of general relativity.  The normal mode frequencies are determined
principally by the strength of the tangled field and bulk stellar
properties, with realistic stellar structure and general relativity
coming in as secondary effects. In \S 5 of the supplementary
materials \citep{lv16s}, we show that realistic
structure has a relatively small effect on the normal mode spectrum
for an isotropic tangle. Our approach, therefore, is to proceed with a
very simple model that elucidates the consequences of a tangled
field. We do not expect refinements of the simple model given here to
alter our chief conclusions. 

We treat the magnetic field as consisting of a smooth dipolar
contribution $\Bbf_d$, plus a tangled component $\Bbf_t$ that
stabilizes the field. At location $\rbf$, the field is
\begin{equation}
\Bbf(\rbf)=\Bbf_d(\rbf)+\Bbf_t(\rbf).
\end{equation}
We assume that $\Bbf_t(\rbf)$ averages to nearly zero over a dimension
of order the stellar radius or smaller, that is,
$\ev{\Bbf_t(\rbf)}\simeq 0$ where $\ev{...}$ denotes a volume
average; see 
the supplementary materials \citep{lv16s} for details.
The magnetic energy density in the tangle is
$\ev{B_t^2}/8\pi$.  We define a dimensionless measure of the strength
of the tangle as the ratio of the energy density in the tangle to that
in the dipolar component:
\begin{equation}
b_t^2\equiv \frac{\ev{B_t^2}}{B_d^2}.
\end{equation}
We regard
the magnetic tangle as approximately isotropic, and the dominant
source of magnetic stress, so that $b^2_t>>1$, as supported by the
simulations and arguments of \cite{braithwaite09}. In this limit, the
fluid core acquires an effective shear modulus given by: 
\begin{equation}
\mu_B\equiv \frac{\ev{B_t^2}}{4\pi}; 
\end{equation}
We limit the analysis to low-frequency QPOs, as the approximation of
an isotropic tangle could break down at high frequencies.

Using realistic structure calculations (see \S \ref{crust}), the volume
averaged-shear modulus of the crust is $\bar{\mu}_c\simeq 3\times
10^{29}$ \ergcc. The magnetic rigidity of the tangle will dominate 
the material rigidity of the crust when 
\begin{equation}
\frac{\bar{\mu}_c}{\mu_B}<<1 \longrightarrow
b_t^2 >> \left(\frac{\bar{\mu}_c}{3\times 10^{29}\mbox{ \gcc}}\right)
\left(\frac{B_d}{10^{15}\mbox{ G}}\right)^{-2}, 
\label{crust_criterion}
\end{equation} 
and so the crust is dynamically negligible in the same limit
(coincidentally) that the isotropic tangle becomes the dominant form
of magnetic stress for a typical magnetar dipole field of $10^{15}$ G;
we quantify the small effect of the crust in \S \ref{crust}.  We
ignore the crust in our simple model, and treat the star as a
self-gravitating, constant-density, magnetized fluid whose torsional
normal modes are determined by the isotropic stresses of the tangled
field. General relativity is included as a redshift factor that
reduces the oscillation frequencies observed at infinity by about
20\%. Electrical resistivity is negligible for the modes of interest,
and we work with ideal MHD.  Our normal-mode analysis is restricted
entirely to toroidal modes. The equations of motion are derived
using a mean-field formalism in \S 1 of the supplementary materials 
\citep{lv16s}; the equation of motion for small displacements of the
fluid $\ubf$ is
\begin{equation}
c_t^2\nabla^2\ubf+\omega^2\ubf=0, 
\label{waveeqn}
\end{equation}
where $c_t\equiv(\ev{B_t^2}/4\pi x_p\rho)^{1/2}$ is the Alfve\'n wave
speed through the tangle, $\rho$ is the mass density, $x_p\simeq 0.1$
is the proton mass fraction fraction, and $\omega$ is an
eigenfrequency. In evaluating $c_t$, we have assumed that the neutrons
are superfluid. If the protons are normal, the neutrons do not scatter
with the protons (ignoring scattering processes with the vortices of
the neutron superfluid). 
If the protons are superconducting, the neutron
fluid is entrained by the proton fluid, but this effect is
negligible \citep{ch06}. In either case, the dynamical mass density is
essentially $x_p\rho$. 

At this point we have reduced the normal mode problem to that of an
elastic sphere of constant density and rigidity. 
Torsional modes have the form $\ubf = u_\phi(r,\theta) \hat{\phi}$
in spherical coordinates $(r,\theta,\phi)$ with the origin at the
center of the star.
The solutions to eq. (\ref{waveeqn}) are (see \S 3 of the
supplementary materials \citep{lv16s} for
further details):
\begin{equation}
u_\phi(r)=A j_l(kr)\,\frac{dP_l(\theta)}{d\theta},
\label{sol}
\end{equation}
where $k\equiv\omega/c_t$ and 
$A$ is normalization. The eigenfunctions and associated
eigenfrequencies are determined by the boundary condition that the
traction vanish at the stellar surface: 
\begin{equation}
\left[\frac{dj_l}{dr}-\frac{j_l}{r}\right]_{r=R}=0.
\label{bc_isotropic}
\end{equation}

For each value of $l$, eq. (\ref{bc_isotropic}) has solutions
$x_{ln}\equiv k_{ln} R$, where $n=0,1,2...$, the overtone number,
gives the number of nodes in $j_l(kr)$. The eigenfrequencies are
\begin{equation}
\omega_{ln}=z
\left(\frac{\ev{B_t^2}R}{3x_pM}\right)^{1/2} x_{ln},
\end{equation}
where a redshift factor $z\equiv(1-R_s(M)/R)^{1/2}$ has been
introduced; $R_s$ is the Schwarzchild radius.\footnote{For $l=1$, 
eq. (\ref{waveeqn}) has a
solution $u_\phi =A\, r\, dP_1/d\theta= A\,r$ for $\omega=k=0$.  This solution,
which we label $n=0$, corresponds to rigid-body rotation, and so is of
no physical significance to the mode problem we are addressing. All
other modes have zero angular momentum.} In terms of fiducial
values
\begin{equation}
\nu_{ln}(\mbox{Hz})=\frac{\omega_{ln}}{2\pi}=
4.3\, \left(\frac{z}{0.77}\right)
\left(\frac{R}{10\mbox{ km}}\right)^{1/2}
\left(\frac{M}{1.4M_\odot}\right)^{-1/2}
\left(\frac{x_p}{0.1}\right)^{-1/2}
\left(\frac{\ev{B_t^2}^{1/2}}{10^{15}\mbox{ G}}\right)
\, x_{ln}\mbox{ Hz}.
\label{nu_isotropic}
\end{equation}

The energy in a torsional mode $(l,n)$ is 
\begin{equation}
E_{ln}=\frac{1}{2} \int d^3r\, x_p \rho\, \omega^2 (u_\phi^{ln})^2.
\label{E}
\end{equation}
For the purpose of comparing the energies
of different normal modes of the same amplitude, we choose the
normalization $A$ in eq. (\ref{sol}) so that
\begin{equation}
\bar{u}^2=\int_{4\pi}d\Omega\, u_\phi(R)^2, 
\label{norm}
\end{equation}
where $\bar{u}^2$ is the square of the mode amplitude averaged over
the stellar surface. We evaluate the energy in a mode $(l,n)$,
normalized by the $l=2$, $n=0$ mode energy, with $\bar{u}^2$ set equal
for both modes. 

In the second row of Table \ref{nu_SGR1900_values}, we give example
frequencies for the fundamental of each $l$ for $M=1.4M_\odot$, $R=10$
km, and $x_p=0.1$. Numbers in parenthesis give the normalized energy
in the mode. By tuning the parameter
$b_t^2=\ev{B_t^2}/B_d^2$ to 14.4, the spectrum of the fundamentals for
each $l$ agrees with the QPOs seen in SGR 1900. A more accurate
analysis given in \S 4 of the supplementary materials \citep{lv16s}
changes $b_t^2$ slightly to 13.3.  Taking $B_d$ equal to
the inferred dipole field for SGR 1900 of $7\times 10^{14}$ G, this
value of $b_t^2$ corresponds to $\ev{B_t^2}^{1/2}=2.7\times 10^{15}$
G. We also show some of the eigenfrequencies for the first three
overtones. The overtones begin at higher frequencies; they also
require more energy to excite to the same root-mean square amplitude
(eq. \ref{norm}) than the fundamental modes and so are energetically
suppressed. To the extent that the surface amplitude of a mode
determines its observability through variations of magnetospheric 
emission, the overtones might not be as important as the fundamental
modes. The amplitude of a given mode depends on the excitation
process, and we note that overtones could prove relevant in a more
detailed treatment that addresses the initial-value problem of mode
excitation; we discuss this issue further below. 

The fundamental modes are
nearly evenly spaced in $l$, with frequencies given by
\begin{equation}
\nu_l(\mbox{Hz})\simeq 0.5\, l\, \nu_2,
\label{sequence}
\end{equation}
where $\nu_2$ is the frequency of the $l=2$ fundamental. The mode
spacing is about half of $\nu_2$. From eq. (\ref{nu_isotropic}), the
lowest-frequency mode and the mode spacing both scale 
as 
$z(M/R)^{-1/2}$.  The
observed QPO spacing follows eq. (\ref{sequence}), though only four of
the 12 modes in the range $2\le l\le 12$ are seen; we discuss this
point further below.

The highest fundamental frequency given in Table
\ref{nu_SGR1900_values} is 156 Hz for $l=12$. The wavelength of this
mode is $\simeq 0.5R$. Hence, for modes in the 28 to 156 Hz range, the
approximation of an isotropic tangle is required to hold over stellar
dimensions, as supported by the simulations of \cite{braithwaite09}.

This model of a tangle that dominates the dipole field does not
apply to SGR 1806. If we attempt to
explain the lowest-frequency QPO observed (18 Hz) as an $l=2$
fundamental mode for the inferred dipole field strength of $2\times
10^{15}$, eq. (\ref{nu_isotropic}) gives $b_t^2\simeq 0.5$, which is
inconsistent with the approximation of a strong, nearly isotropic
tangle. For this case, the magnetic stress of the smooth field must be
included. This problem is solved in \S 4 in the supplementary
materials \citep{lv16s}. A match to the 18 Hz QPO implies
$b_t^2=0.17$. The spectrum is very dense with a spacing of about two
Hz. If the dipole field has been overestimated by a factor several for
this object, not implausible, then the predicted spectrum is much less
dense, and more similar to SGR 1900. For example, taking $B_d$ equal
to 0.62 of the reported value, a value of $b_t^2=1.0$ matches the 18
Hz QPO, and the predicted spectrum is less dense, with a spacing of
about 7 Hz in the fundamentals.

\section{Effects of the crust are negligible}

\label{crust}

So far we have ignored the crust under the assumption that magnetic
stresses throughout the star dominate material stresses in the
crust. Here we show that crust rigidity increases the
eigenfrequencies calculated above for SGR 1900 by 3\% or less. A more
detailed treatment is given in \S 4 of the supplementary materials
\citep{lv16s}.

To estimate the effects of the crust, we use a two-zone crust plus
core model, assuming a nearly isotropic tangle ($b_t^2>>1$). The core
has constant density $\rho$ and constant effective shear modulus
$\mu_B$. The crust has inner radius $R_c$, outer radius $R$, thickness
$\Delta R$, average density $\bar{\rho}_c$, average material shear
modulus $\bar{\mu}_c$, and average effective shear modulus
$\bar{\mu}_{\rm crust}=\bar{\mu}_c+\mu_B$. \citet{chamel05,chamel12}
finds that the neutron fluid is largely entrained by ions in the inner
crust; in evaluating the wave speed in the crust, we use the total
mass density, so that the wave propagation speed in the crust is
$\bar{c}_{\rm crust}=\sqrt{\bar{\mu}_{\rm crust}/\bar{\rho}_c}$. In
the core the propagation speed is $c_t=\sqrt{\mu_B/x_p\rho}$, where 
we take the proton mass fraction to be $x_p=0.1$ (see discussion after
equation \ref{waveeqn}).

In the core, the solution to the mode
problem is $u_{\rm core}=j_l(kr)$. The solution in the crust is 
$u_{\rm crust}=aj_l(k^\prime r)+bn_l(k^\prime r)$, where $n_l$ are spherical
Neumann functions, and $a$ and $b$ are constants. The wavenumbers are
related through $\omega=c_tk=\bar{c}_{\rm crust}k^\prime$. 
The boundary conditions are continuity in value and
traction at $r=R_c$, and vanishing traction at $r=R$:
\begin{equation}
u_{\rm core}(R_c)=u_{\rm crust}(R_c),
\end{equation}
\begin{equation}
\mu_B\left[\frac{du_{\rm core}}{dr}-\frac{u_{\rm core}}{r}
\right]_{r=R_c}=
\bar{\mu}_{\rm crust}\left[\frac{du_{\rm crust}}{dr}-\frac{u_{\rm crust}}{r}
\right]_{r=R_c},
\end{equation}
\begin{equation}
\left[\frac{du_{\rm crust}}{dr}-\frac{u_{\rm crust}}{r}
\right]_{r=R}=0.
\end{equation}
For $\bar{c}_{\rm crust}$ and $\bar{\mu}_{\rm crust}$, we use
volume-averaged quantities obtained in the following way. The shear
modulus in the crust as a function of density, ignoring magnetic
effects, is in cgs units
\citep{strohmayer_etal91} 
\begin{equation}
\mu_c=0.1194\,\frac{n_i(Ze)^2}{a}, 
\end{equation}
where $n_i$ is the number density of ions of charge $Ze$, and $a$ is the
Wigner-Seitz cell radius given by $n_i4\pi a^3/3=1$.
For the composition of the inner crust, we use the results
of
\citet{dh01}, conveniently expressed analytically
by \citet{hp04}. 
We solve for crust structure using the Newtonian equation for
hydrostatic equilibrium for a neutron star of 1.4 $M_\odot$. The
volume-averaged density and shear modulus in the crust are 
$\bar{\rho}_c=0.06\rho$ and $\bar{\mu}_c=2.5\times 10^{29}$ \ergcc,
respectively. The crust thickness is $\Delta
R=0.1R$. We take $\ev{B_t^2}^{1/2}=2.7\times 10^{15}$ G assumed
above for SGR 1900, corresponding to $\mu_B=5.8\times 10^{29}$
\ergcc.

To evaluate the effects of the crust, we solve the two-zone model for
a particular eigenmode first taking $\bar{\mu}_{\rm
crust}=\bar{\mu}_c+\mu_B$, then taking $\bar{\mu}_{\rm crust}=\mu_B$,
and evaluating the difference between the two eigenfrequencies. We
find that the finite shear modulus of the crust increases the
frequencies of the fundamental normal modes by only about 3\% for
$l=2,3,4$, and 2\% for $l=12$. We have confirmed that the effects are
even smaller for overtones. The crust is nearly dynamically irrelevant
in this limit of a strong magnetic tangle, and neglect of the crust is
a good approximation.

\section{Discussion and Conclusions}

\label{conclusions}

Theoretical study of magnetar QPOs is seriously hampered by the fact
that we do not know the detailed magnetic structure within a
magnetar. Even if the magnetic structure were known or is
assumed, solution of the full problem is very difficult. Motivated by
stability considerations and observational evidence, we have argued
that the magnetic field should be highly tangled; {\em significant tangling
of the magnetic field is needed to stabilize the linked poloidal and
toroidal fields} \citep{bn06,braithwaite08}.  We have shown with a
simple model that a highly-tangled field, under the assumption that
the tangle is approximately isotropic, supports normal modes with
frequencies consistent with the QPOs observed in SGR 1900. In
comparison with data, we have obtained a rough measurement of the
ratio of the energy density in the tangle to that in the dipolar field
of $\sim 14$. Given the approximations we have made, this model should
not be taken as quantitatively very accurate, but what is significant
is that the assumption of a strong magnetic tangle leads naturally to a
normal mode spectrum with frequencies that lie in the range of
observed QPOs. Our model predicts about three times as many modes
below 160 Hz than are observed. That the normal mode spectrum is more
dense than the observed QPO spectrum is crucial, for if the
opposite were true, we would be forced to abandon the model as
unviable.

The main point that we would like to emphasize is that field tangling has 
important effects that cannot be ignored in the QPO problem, and that
the magnetic tangle is likely to be the principal factor that
determines the normal-mode spectrum. Use of dipolar magnetic
geometries and their variants is not adequate, and we point out that
such magnetic geometries are unstable and therefore unphysical starting points
for study of normal modes of neutron stars. One simplifying feature of
magnetic tangling is that for a tangle of magnetar-scale strength,
the crust becomes dynamically unimportant. 

We conclude by briefly describing several directions of future
research that we consider to be interesting and important. One issue
is the prediction that the normal-mode spectrum is denser than the
observed QPO spectrum.  In general, any excitation mechanism will give
preferential mode excitation. Determination of which modes are excited
requires solution of the initial-value problem.  Also, it is unknown
at present if the QPO emission is beamed or not. If the emission has a
beaming fraction less than unity, only that fraction of modes would be
potentially observable. We are currently studying the mode excitation
problem; we find that excitation in a localized region of the star can
lead to excitation of separate groups of modes.

Further development of the model into a quantitative tool will require
inclusion of realistic stellar structure and treatment of the case of
comparable energy densities in the magnetic tangle and the dipolar
component. For SGR 1806, the energy densities in the
tangle and the dipolar component appear to be comparable within our
interpretation; see \S 4 of the supplementary materials
\citep{lv16s}. In this case, the spectrum of normal modes could be very
dense, with a mode spacing of about 2-7 Hz. 
If the spectrum is so dense that it begins to
assume the qualities of a quasi-continuum, resonant absorption might
be important as has been considered in previous work with smooth field
geometries. We will study this interesting problem in future work.
The magnetic field that occurs in nature may not be a
nearly isotropic tangle as we have assumed. It would be interesting to
explore how spatial variations in the magnetic tangle affect the
spectrum of normal modes. 

\begin{acknowledgments}

We thank M. Gabler, D. Huppenkothen, Y. Levin, and A. Watts for very
helpful discussions, and the anonymous referee for useful criticism. This
work was supported by NSF Award AST-1211391 and NASA Award NNX12AF88G.

\end{acknowledgments}


\afterpage{%
    \clearpage
    \thispagestyle{empty}
    \begin{landscape}
        \centering 

\begin{table}
\begin{tabular}{lllllllllllll}
\hline
\hline
$l$ & 1  & 2 & 3 & 4 & 5 & 6 & 7 & 8 & 9 & 10 & 11 & 12 \\
\hline
$n=0$ & - & {\bf 28}(1) & 43(2) & {\bf 57}(4) & 70(5) & {\bf 83}(7) & 95(8) & 108(10) & 120(12) & 132(14) & 144(16)
& {\bf 156}(18) \\
$\nu_{\rm obs} $ &  & $28\pm 2$ & & $53\pm 5$ & & 84 & & & & & & $155\pm 6$  \\ 
\hline
$n=1$ & 102(37) & 80(21) & 95(27) & 109(34) & 123(41) & 136(48) &
150(55) & 163(63) & & & &  \\
$n=2$ & 138(67) & 118(47) & 133(58) & 148(69) &  162(81) & & & & & & & \\
$n=3$ & & 154(82) & & & & & & & & & & \\
\hline
\hline
\end{tabular}
\caption{Example eigenfrequencies for SGR 1900 with
$b_t^2=14.4$, corresponding to $\ev{B_t^2}^{1/2}=2.7\times 10^{15}$
G. All frequencies are in Hz. Numbers in parentheses indicate the
factor of energy required to excite the mode to the same amplitude as
the $l=2$ fundamental. 
Numbers in boldface lie close to observed
QPOs (third row), and represent plausible mode
identifications. The fourth, fifth, and sixth rows show the $n=1$,
$n=2$, and $n=3$ overtones. The overtones begin at higher frequencies
than do the fundamentals, and require more energy to excite to the
same amplitude. We only list frequencies below 163 Hz.  }
\label{nu_SGR1900_values}
\end{table}

   \end{landscape}
    \clearpage
}

\newpage

\noindent
{\bf \large Supplement to ``TORSIONAL OSCILLATIONS OF A MAGNETAR WITH
A 
TANGLED MAGNETIC FIELD'' (ApJL, {\sl in press}) }

\setcounter{equation}{0}
\setcounter{section}{0}
\setcounter{page}{1}
\setcounter{figure}{0}
\setcounter{table}{0}

\bigskip\bigskip

\centerline{\bf\large ABSTRACT}

\bigskip

\parbox{15.cm}{

\noindent
We introduce a mean-field formalism with which to study the
oscillations of a neutron star or other stellar object with a dipolar
magnetic field plus a topologically distinct isotropic magnetic tangle
that stabilizes the field configuration. In terms of the ratio of the
energy density of the tangled field to that in the dipolar component
$b_t^2$, we obtain separable equations for the eigenfunctions and
eigenfrequencies for a star of uniform density. We show that finite
$b_t^2$ breaks the Alfv\'en continuum that is supported by the dipolar
component of the field into discrete normal modes, and we quantify the
splitting. Assuming the dipole fields estimated from spin down of
$7\times 10^{14}$ G in SGR 1900+14 and $2\times 10^{15}$ G in SGR
1806-20, and tuning $b_t^2$ to match the lowest-frequency
quasi-periodic oscillations observed to accompany flares in these
objects, we infer $b_t^2=13.3$ and 0.17, respectively. The predicted
spectrum is about three times denser than observed in SGR 1900+14, and
much denser in SGR 1806-20. If the dipole field of SGR 1806-20 has
been overestimated, the spectrum of normal modes will be less dense. We
show that for $b_t^2>>1$, as stability considerations suggest, the
crust is nearly dynamically irrelevant. Moreover, density
stratification generally shifts the eigenfrequencies up by about 20\%,
a change that can be mostly subsumed in a uniform-density model with a
slightly different choice of stellar radius or mass.

}

\section{Equations of Motion}

\label{eom}

The Maxwell stress tensor for
matter permeated by a field $\Bbf$ is 
\begin{equation}
T_{ij}=\frac{1}{4\pi}\left[B_iB_j-\frac{1}{2}B^2\delta_{ij}\right]
\label{T}
\end{equation}
Let the matter be displaced by $\ubf$. 
For a perfect conductor, perturbations in the field satisfy
\begin{equation}
\delta\Bbf=\nabla\times(\ubf\times\Bbf).
\label{flux_freezing}
\end{equation}
We
specialize to shear perturbations, so that $\nabla\cdot\ubf=0$, for
which equation (\ref{flux_freezing}) becomes
\begin{equation}
\delta\Bbf=(\Bbf\cdot\nabla)\ubf-(\ubf\cdot\nabla)\Bbf.
\end{equation}
For a displacement $\ubf$, the stress tensor is perturbed by
\begin{equation}
\delta T_{ij}=\frac{1}{4\pi}\left[
B_jB_k\nabla_k u_i 
-B_ju_k\nabla_k B_i 
-\frac{1}{2}\delta_{ij}B_kB_l\nabla_lu_k
+\frac{1}{2}\delta_{ij}B_k u_l\nabla_l B_k\right] 
+\mu\nabla_iu_j+\mbox{ transpose}, 
\label{dT}
\end{equation}
where repeated indices are summed, $B_i$ denotes a component
of the unperturbed field, and $\mu$ is the shear modulus of the
crust. 

We decompose the magnetic field at location $\rbf$ in the star as 
\begin{equation}
\Bbf(\rbf)=\Bbf_d(\rbf)+\Bbf_t(\rbf), 
\end{equation}
where $\Bbf_d(\rbf)$ denotes the local dipolar contribution, and
$\Bbf_t(\rbf)$ denotes a topologically-distinct magnetic tangle that
gives a globally-stable field.  We assume that the field is tangled
for length scales smaller than some length scale $l_t$, and that $l_t$ is
smaller than the wavelengths of the eigenfrequencies of
interest. We also assume that over length scales that exceed $l_t$,
the magnetic tangle is approximately isotropic. 
For calculational simplicity, we take $\Bbf_d$ to be
constant. We denote volume averages over $l_t^3$ as $\ev{...}$. 

We assume that different components of the tangled field are
uncorrelated on average. Under this assumption, the tangled field can
contribute only isotropic stress over length scales above $l_t$, so
that
\begin{equation}
\ev{B_iB_j}=B^d_iB^d_j +\ev{B_t^2}\delta_{ij},
\label{dc1}
\end{equation}
where $\ev{B_t^2}$ is a constant. 

To treat the tangled field's contribution to the stress, we average
the perturbed stress tensor of equation (\ref{dT}), using equation (\ref{dc1}),
 to obtain
\begin{equation}
\langle\delta T_{ij}\rangle=
\frac{1}{4\pi}\left[
\ev{B_jB_k}\nabla_k u_i -u_k\ev{B_j\nabla_kB_i}
-\frac{1}{2}\delta_{ij}\ev{B_kB_l}\nabla_l u_k +
\frac{1}{2}\delta_{ij}u_l\ev{B_k\nabla_lB_k}\right]
+\mu\nabla_i u_j + \mbox{ transpose}, 
\label{dT1}
\end{equation}
where $u_i$ now denotes a component of the displacement field averaged
over $l_t$, that is, $u_i\equiv\ev{u_i}$. 

If different components of the tangled field are uncorrelated over
$l_t^3$, one component will also be uncorrelated with the gradient
of a different component, that is, 
\begin{equation}
\ev{B_i\nabla_k B_j}=\ev{(B^d_i+B^t_i)\nabla_k (B^d_j+B^t_j)}=
\ev{B^t_i\nabla_k B^t_j}=0 \qquad i\ne j.
\label{dc2}
\end{equation}
Since the tangled field varies over length scales smaller than $l_t$, 
a component of the tangled field will also be uncorrelated with the
gradient of the same component, so that
\begin{equation}
\ev{B^t_i\nabla_k B^t_i}=0, 
\label{dc3}
\end{equation}
which applies component by component, and therefore also in
summation. 
Using equations (\ref{dc1}), (\ref{dc2}), (\ref{dc3}), and
$\nabla\cdot\ubf=0$, equation (\ref{dT1}) becomes
\begin{equation}
\ev{\delta T_{ij}}=
\frac{1}{4\pi}\left(
B_j^d B_k^d\nabla_k u_i + B_i^d B_k^d\nabla_k u_j
-\delta_{ij}B_k^d B_l^d \nabla_l u_k\right)
+\left(\frac{1}{4\pi}\ev{B_t^2}+\mu\right)
\left(\nabla_i u_j +\nabla_j u_i\right).
\label{dT_final}
\end{equation}
The tangled field gives the fluid an effective shear modulus of
$\ev{B_t^2}/4\pi$, and enhances the rigidity of the solid.
In \S \ref{stratification}, we show that crust rigidity has a small
effect on the torsional eigenfrequencies for $b_t^2>>1$. We ignore the
rigid crust until \S \ref{stratification}.

We will be interested in modes with wavelengths greater than $l_t$, for
which the equation of motion is 
\begin{equation}
\rho_d\frac{\partial^2 u_j}{\partial t^2}=\nabla_i \ev{\delta T_{ij}}, 
\label{eom_gen}
\end{equation}
where $\rho_d$ is the dynamical mass density of matter that is frozen
to the magnetic field. If the protons are normal, the superfluid
neutrons do not scatter with the protons. If the protons are
superconducting, the neutron fluid is only slightly entrained by the
proton fluid \citep{ch06a}. In either
case, the dynamical mass density is effectively $\rho_d=x_p\rho$,
where $\rho$ is the total mass density, and $x_p\simeq 0.1$ is the
mass fraction in protons.

We neglect coupling of the stellar surface to
the magnetosphere, and treat the surface as a free
boundary with zero traction, thus ignoring momentum flow into the
magnetosphere. Under this assumption, the traction at the surface vanishes:
\begin{equation}
\hat{r}_i\ev{\delta T_{ij}}=0,
\label{eom1}
\end{equation}
where $\hat{r}$ is the unit vector normal to the stellar surface. 

We now consider 
a uniform star of density $\rho=3M/4\pi R^3$, where
$M$ and $R$ are the stellar mass and radius, and take
$\Bbf_d=\hat{z}B_d$ where $B_d$ is constant.  Equations (\ref{dT_final})
and (\ref{eom_gen}) give
\begin{equation}
c_d^2\frac{d^2\ubf}{dz^2}
-c_d^2\nabla\frac{d u_z}{dz}+
c_t^2\nabla^2\ubf+\omega^2 \ubf=0,
\label{eom_vector}
\end{equation}
where $c_d^2\equiv B_d^2/(4\pi\rho_d)$ and $c_t^2\equiv
\ev{B_t^2}/(4\pi\rho_d)$; $c_d$ is the speed of Alfv\'en waves
supported by the dipole field, and $c_t$ is the speed of transverse
waves supported by the isotropic stress of the tangled field. We show
in \S \ref{stratification} that realistic stellar structure shifts the
eigenfrequencies by about 20\% or less, a shift that can be mostly
subsumed in the constant-density model with slightly different choices
of stellar radius or mass. The constant-density model is a good
approximation to a realistic star, and affords straightforward
analysis of the eigenvalue problem.

\section{The Alfv\'en Continuum}

\label{continuum}

For finite $\Bbf_d$ and $\Bbf_t=0$, there exists a
continuum of axial modes \citep{levin07a}, 
given by equation (\ref{eom_vector}) 
\begin{equation}
c_d^2\frac{d^2\ubf}{dz^2}-c_d^2\nabla\frac{d u_z}{dz}+\omega^2\ubf=0, 
\end{equation}
In cylindrical coordinates ($s,\phi,z$), axial modes are given by
$\ubf=u_\phi\,\hat{\phi}$. 
For a constant field, and no crust, field
lines have a continuous range of lengths between zero and $2R$,
determined by the cylindrical radius $s$. 
Within the approximation of ideal
MHD, fluid elements at different cylindrical radii cannot exchange
momentum. At given $s$, the length of a field line is
$2\sqrt{R^2-s^2}$. The 
solutions have even parity ($\cos kz$) and odd parity ($\sin kz$).  The
requirement that the traction vanish at the stellar surface gives the
spectrum
\begin{equation} \label{cont_even}
\nu_n=\frac{n}{2}\frac{c_d}{\sqrt{R^2-s^2}} 
\qquad\mbox{even parity}
\end{equation}
\begin{equation} \label{cont_odd}
\nu_n=\frac{(2 n+1)}{4}\frac{c_d}{\sqrt{R^2-s^2}}  \qquad\mbox{odd parity.}
\end{equation}
where $n$ is an integer, beginning at zero for the odd-parity
modes, and $\nu=\omega/2\pi$.  Because $s$ is a
continuous variable, for every $n$ there is a continuous spectrum of
modes for this simple magnetic geometry.  An infinite sequence of
continua begins at $\sim 7 (2n+1) \,B^d_{15}$ Hz for odd-parity modes
and $ \sim 7 (2 n)
\,B^d_{15}$ Hz for the even-parity modes, where $B^d_{15}\equiv
B_d/(10^{15}\mbox{ G})$.  The spectrum begins at $\sim
7\,B^d_{15}$ Hz; there is also a zero-frequency mode corresponding to
rigid-body rotation.  The same conclusion holds for more general
axisymmetric field geometries, though certain geometries give gaps in
the continuum.

\section{Isotropic Model}

\label{isotropic_model}

We now turn to the opposite extreme of a tangled field that dominates
the stresses, taking $\Bbf_d=0$, and solving the resulting isotropic
problem. This problem provides useful insight into the mode structure
of the more general problem with non-zero $\Bbf_d$ and $\ev{B_t^2}$. 
For this case, equation (\ref{eom_vector}) becomes
\begin{equation}
c_t^2\nabla^2\ubf+\omega^2\ubf=0.
\end{equation}
Subject to the restriction $\nabla\cdot\ubf=0$, the solutions for
spheroidal modes ($u_r=0$), can be
separated as
\begin{equation}
u_\phi=w(r)\frac{\partial}{\partial\theta} Y_{lm}(\theta,\phi){\rm
e}^{i\omega t}
\end{equation}
\begin{equation}
u_\theta=-w(r)\frac{1}{\sin\theta}\frac{\partial}{\partial\phi} 
Y_{lm}(\theta,\phi){\rm
e}^{i\omega t}. 
\end{equation}
The radial function $w(r)$ satisfies Bessel's equation: 
\begin{equation}
\left(\frac{d^2}{dr^2}+\frac{2}{r}\frac{d}{dr}-
\frac{l(l+1)}{r^2}+k^2\right)w(r)=0,
\label{radial_equation}
\end{equation}
where $k\equiv\omega/c_t$. 

The solutions that are bounded at $r=0$ are the spherical Bessel
functions $j_l(kr)$. 
Zero traction at the stellar surface gives
\begin{equation}
\left[\frac{dj_l}{dr}-\frac{j_l}{r}\right]_{r=R}=0.
\label{bc_isotropic1}
\end{equation}
For each value of $l$, equation (\ref{bc_isotropic1}) has solutions
$x_{ln}\equiv k_{ln} R$, where $n=0,1,2...$, the overtone number,
gives the number of nodes in $j_l(kr)$. The eigenfrequencies are
\begin{equation}
\omega_{ln}=\left(1-\frac{R_s(M)}{R}\right)^{1/2}
\left(\frac{\ev{B_t^2}R}{3x_pM}\right)^{1/2} x_{ln}
\end{equation}
where a redshift factor $z\equiv(1-R_s(M)/R)^{1/2}$ has been
introduced; $R_s$ is the Schwarzchild radius. In terms of fiducial
values
\begin{equation}
\nu_{ln}(\mbox{Hz})=\frac{\omega_{ln}}{2\pi}=
4.3\, \left(\frac{z}{0.77}\right)
\left(\frac{R}{10\mbox{ km}}\right)^{1/2}
\left(\frac{M}{1.4M_\odot}\right)^{-1/2}
\left(\frac{x_p}{0.1}\right)^{-1/2}
\left(\frac{\ev{B_t^2}^{1/2}}{10^{15}\mbox{ G}}\right)
\, x_{ln}\mbox{ Hz}.
\label{nu_isotropic1}
\end{equation}
Example eigenfrequencies are given in Table 1 of the accompanying Letter. 

For $l=1$, equation (\ref{bc_isotropic1}) has a solution $w(r)=r$ for
$k=0$.  This
solution corresponds to rigid-body rotation and we label it
$n=0$. This solution is of no physical significance to the mode
problem we are addressing, but we include it in Tables 1-3 for
completeness.

\section{Anisotropic Model}

\label{anisotropic_model}

We now turn to the general problem of non-zero $\Bbf_d$ and
$\ev{B_t^2}$, and show that even a small amount of stress from the
tangled field breaks the Aflv\'en continuum described in \S 2 very
effectively. The normal modes of the system are similar to those found
in the isotropic problem for $\ev{B_t^2}\simeq B_d^2$.

The modes are given by equation (\ref{eom_vector}). 
This vector equation, subject to boundary conditions on the surface of a
sphere, is difficult to solve in general. For illustration, we
specialize to axial modes, $u_\theta=u_r=0$, giving in cylindrical 
coordinates $(s,\phi,z)$:
\begin{equation}
c_d^2\frac{d^2u_\phi}{dz^2}+
c_t^2\nabla^2 u_\phi-c_t^2\frac{u_\phi}{s^2}+\omega^2 u_\phi=0. 
\end{equation}
The third term results from derivatives of $\hat{\phi}$ in
the original vector equation (\ref{eom_vector}).
Defining 
$b_t^2\equiv \ev{B_t^2}/B_d^2$, the ratio of the energy
density in the tangled field to that in the dipole field, the above
equation becomes
\begin{equation}
c_d^2\left[b_t^2\frac{1}{s}\frac{\partial}{\partial s}
\left( s\frac{\partial u_\phi}{\partial s}\right)
+\left(1+b_t^2\right)\frac{\partial^2 u_\phi}{\partial z^2} 
-b_t^2\frac{u_\phi}{s^2}\right]+\omega^2 u_\phi=0\,.
\label{eomt}
\end{equation}
The zero-traction boundary condition (equation \ref{eom1}) 
at the stellar surface becomes 
\begin{equation}
b_t^2\left( s \frac{\partial u_\phi}{\partial s} -u_{\phi} \right)
+\left(1+b_t^2 \right)z\frac{\partial u_\phi}{\partial z}=0.
\label{bct}
\end{equation}
Equation (\ref{eomt}) is solved in the domain $z\geq 0$, and so at
$z=0$ we require
\begin{equation}
  u_\phi =  0\, ,
\end{equation}
for modes with {\it odd} parity about $z=0$, and 
\begin{equation}
  \frac{\partial u_\phi}{\partial z} =  0\, ,
\end{equation}
for modes with {\it even} parity.
The system is separable with a coordinate transformation; the details
are given in Appendix \ref{sov}. We solve equations (\ref{sepu}),
(\ref{sepv}), and (\ref{bcu}) numerically to obtain the normal mode
frequencies. The solutions are given by two quantum numbers: $\kappa$
and the overtone number $n$. $\kappa$ maps smoothly to $l(l+1)$ in the
limit $c_d^2\rightarrow 0$, so we use $l$ and $n$ for convenience in
labeling the modes. We refer to $n=0$ for a given $l$ as the
fundamental for that value of $l$.

\begin{figure*}
\centering
\includegraphics[width=.4\linewidth]{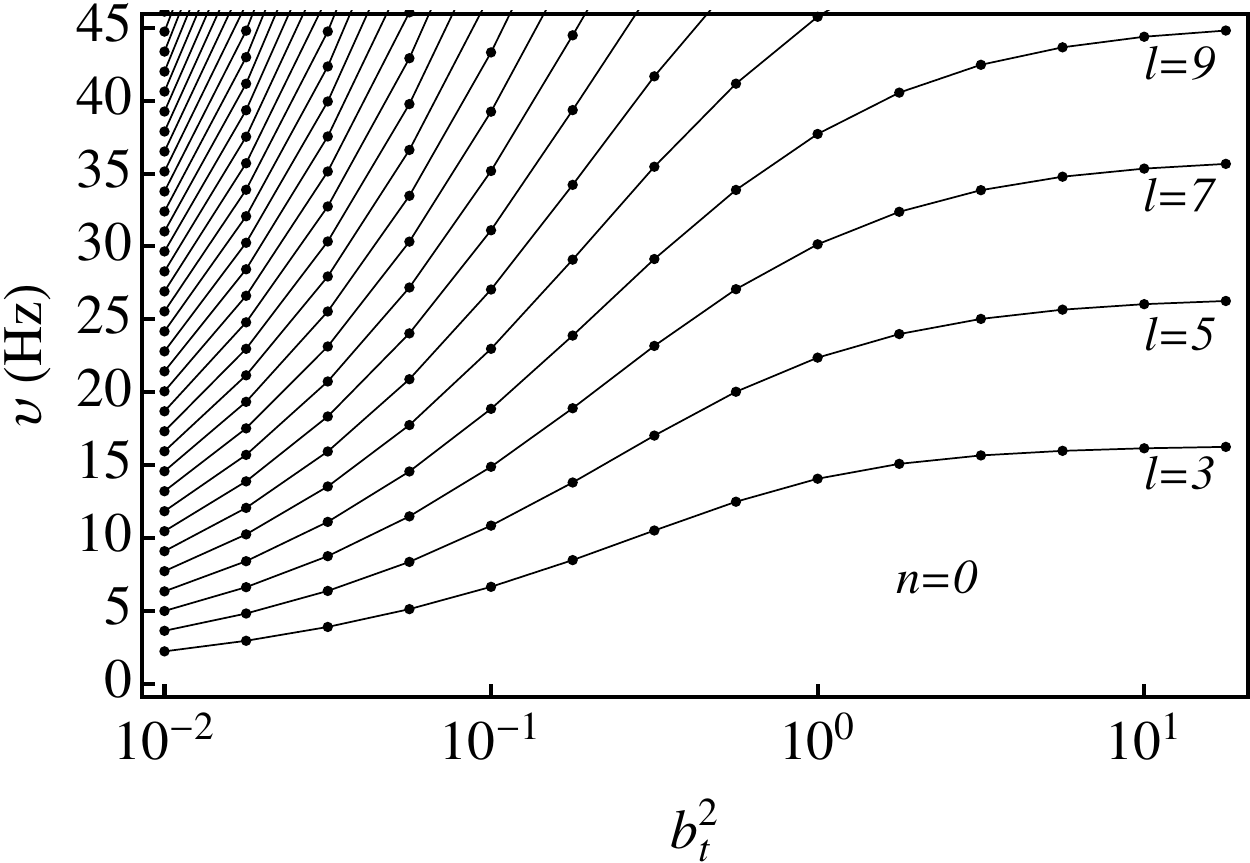} 
\includegraphics[width=.4\linewidth]{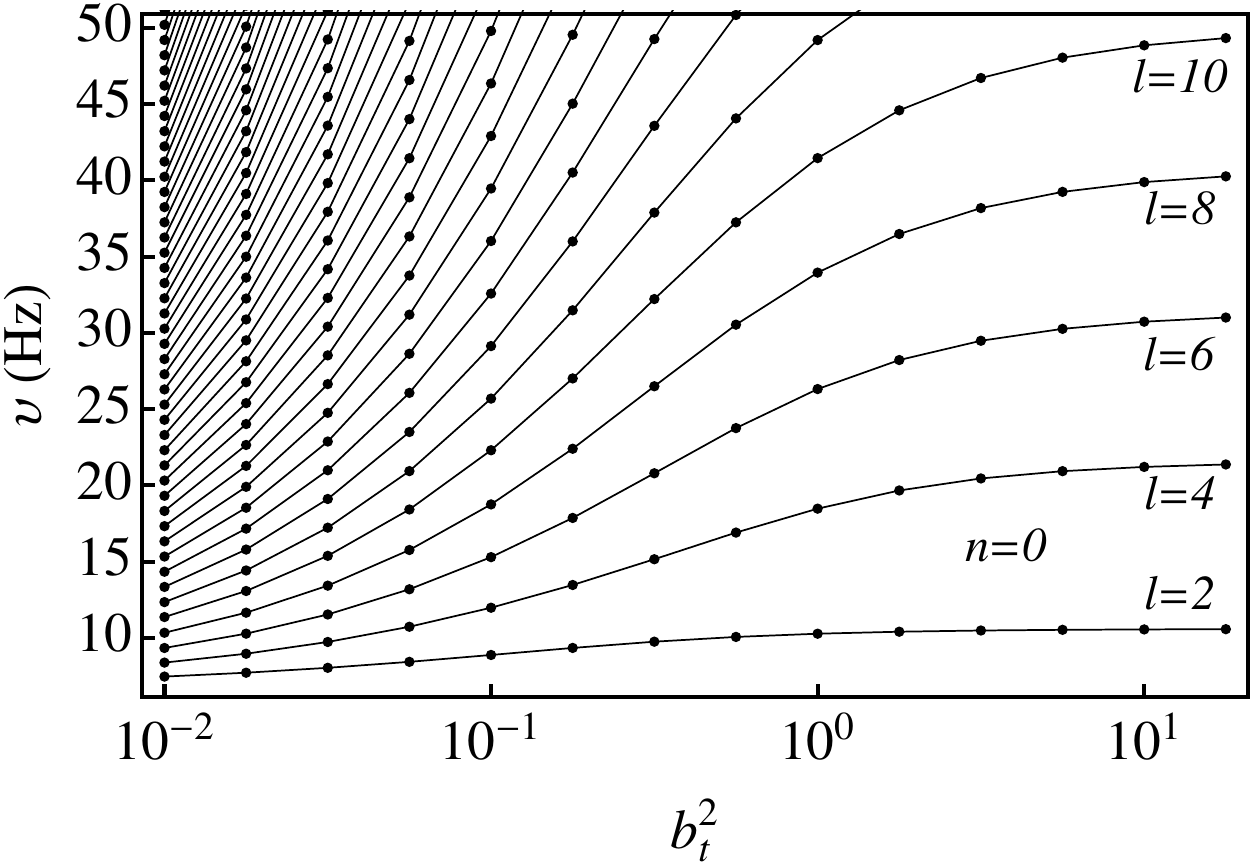} \\
\includegraphics[width=.4\linewidth]{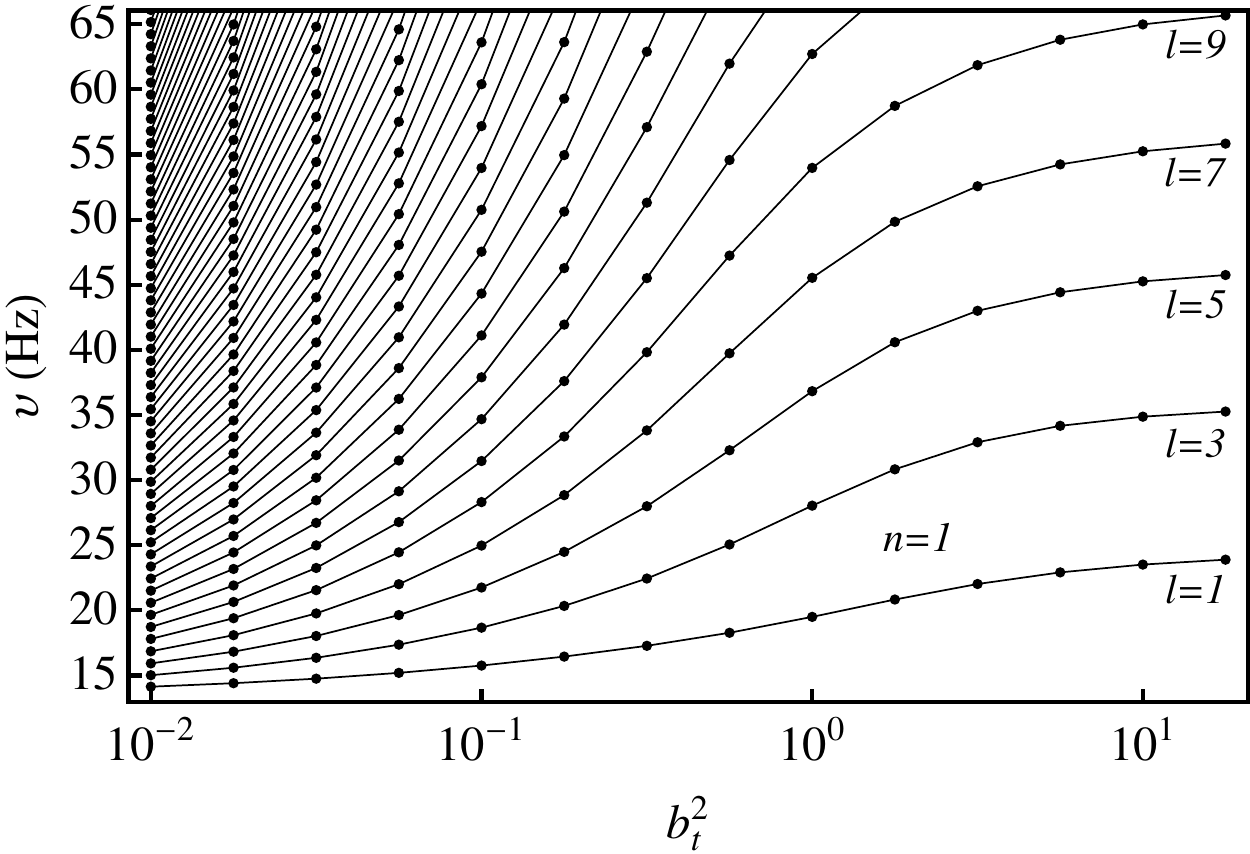} 
\includegraphics[width=.4\linewidth]{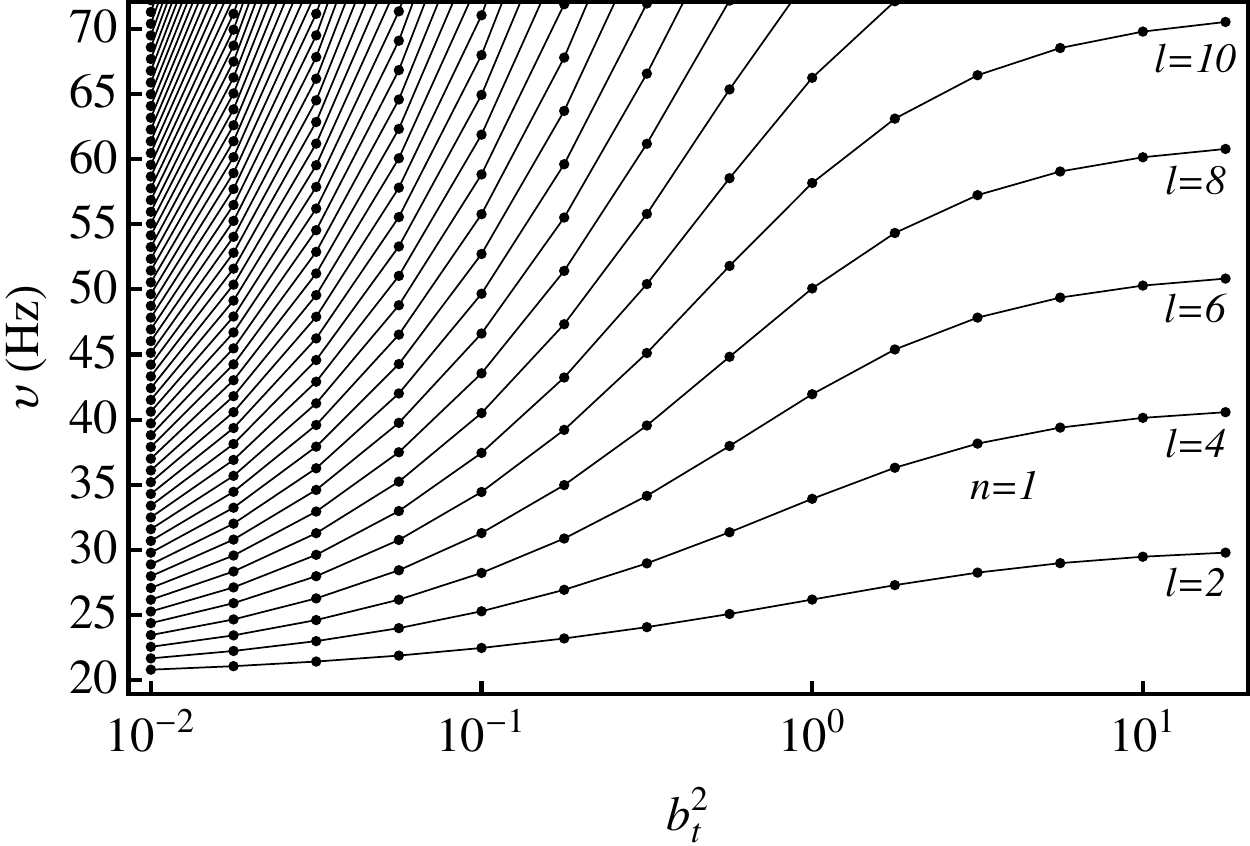}
\caption{Splitting of the Alfv\'en continuum as a tangled field is
added to a smooth field. We have fixed
$(B_d^2+\ev{B_t^2})^{1/2}=B_d(1+b_t^2)^{1/2}$ to $10^{15}$ G to
illustrate the smooth transition from the continuum to the isotropic
tangle. For $l$ odd and $n=0$ (upper left panel), low-frequency
modes exist that go to zero frequency for $b_t^2\rightarrow 0$; see
equation (\ref{l_0_scaling}).}
\label{splitting}
\end{figure*}

The mode structure is shown in Figure \ref{splitting} for constant {\it
total} magnetic energy, with $(B_d^2+\ev{B_t^2})^{1/2}$ fixed at
$10^{15}$ G. 
Modes for which $u_\phi$ has even parity (odd $l$) and odd parity (even $l$) about $z=0$ have been plotted separately.
For $b_t^2>>1$, the solutions to the isotropic problem
(equation \ref{nu_isotropic1}) are recovered.  As $b_t^2$ is reduced, the
modes become more closely spaced, approaching a continuum as
$b_t^2\rightarrow 0$. In the limit $b_t^2\rightarrow 0$ the sequence
of continua begins at $\sim 7(2n+1)$ Hz for odd-parity modes (even
$l$) and $\sim 14 n$ Hz for even-parity modes (odd $l$), in agreement
with the continuum sequences described by equations (\ref{cont_even}) and
(\ref{cont_odd}).  The $n=0$ odd modes approach zero as $b_t^2
\rightarrow 0$ and scale as 
\begin{equation}
\nu=
6.70\, l \, b_t \, \left(\frac{z}{0.77}\right)
\left(\frac{R}{10\mbox{ km}}\right)^{1/2}
\left(\frac{M}{1.4M_\odot}\right)^{-1/2}
\left(\frac{x_p}{0.1}\right)^{-1/2} 
\left(\frac{B_d}{10^{15}\mbox{ G}}\right)
\, \mbox{ Hz}.
\label{l_0_scaling}
\end{equation}
These modes approach rigid-body rotation solutions 
in the $b_t\rightarrow 0$ limit.
Examination of the eigenmodes as $b_t^2\rightarrow 0$ also shows that the oscillation amplitude becomes sharply peaked at a specific value of the cylindrical radius
$s$ and vanishes everywhere else, in agreement with the continuum
solution in \S\ref{continuum}. 

\begin{figure*}
\centering
\includegraphics[width=.48\linewidth]{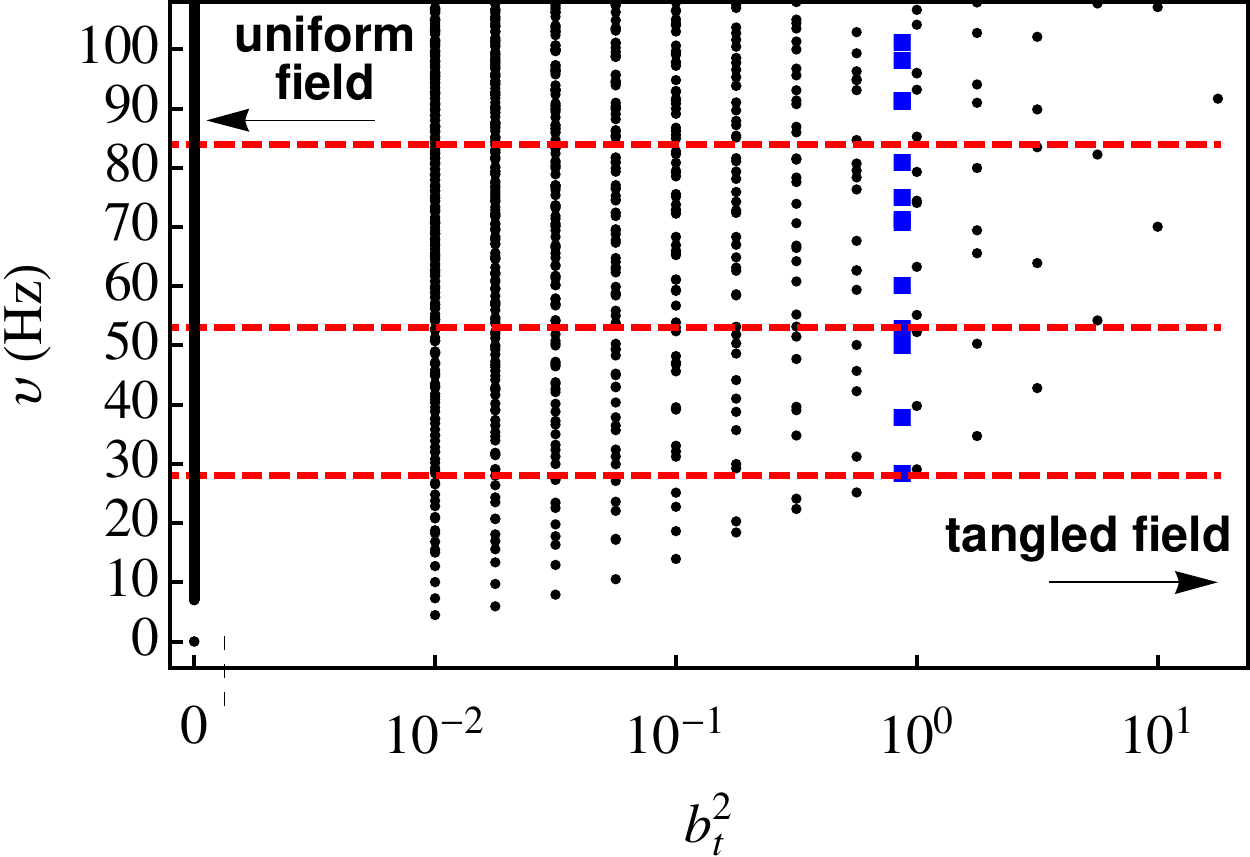} 
\includegraphics[width=.48\linewidth]{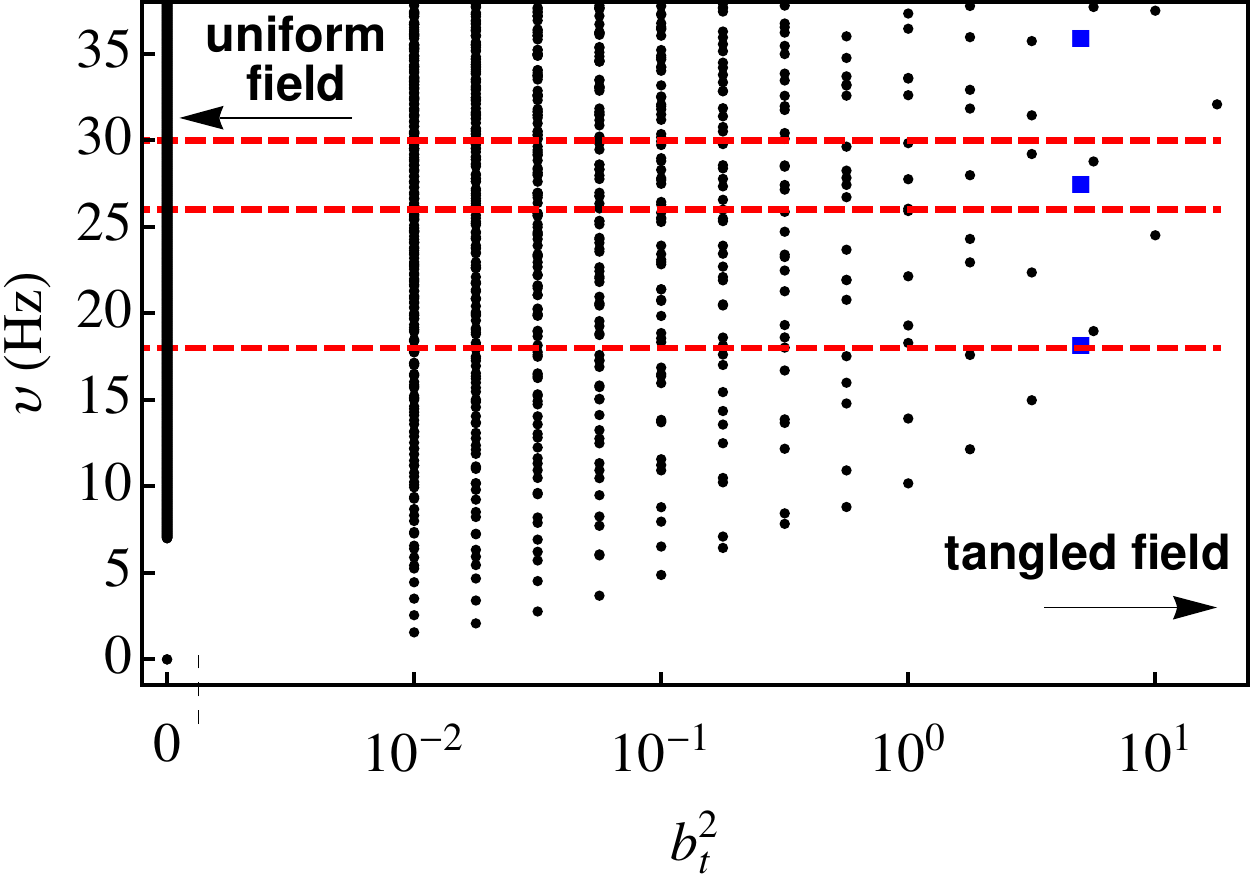} 
\caption{Splitting of the Alfv\'en continuum (the column on the far
left of each figure) as a tangled field is added to a smooth field in
SGR 1900 (left) and SGR 1806 (right). The dipole field strength $B_d$
is fixed to the value inferred from spin-down value: $7\times 10^{14}$
G for SGR 1900 and $2 \times 10^{15}$ G for SGR 1806. Red-dashed lines
correspond to observed QPO frequencies. Solid blue squares give the
eigenfrequencies obtained from choosing the value of $b_t^2$ that
matches the 28 Hz QPO in SGR 1900 and the 18 Hz QPO in SGR 1806. Not
all detected QPOs are shown on the two scales. 
The lowest sequence of points approach the $k=0$ rigid-body
mode, according to equation (\ref{l_0_scaling}).}
\label{dots}
\end{figure*}

The splitting of the Alfv\'en continuum is shown in a different way in
Figure \ref{dots}, where predictions are made for the
eigenfrequencies in SGR 1900 (left) and SGR 1806 (right).  In contrast
with Figure \ref{splitting} where the {\it total} magnetic field was
held constant ($(B_d^2+\ev{B_t^2})^{1/2}=10^{15}$ G), we now fix 
$B_d$ in each magnetar to its observed dipole
spin-down value: $7\times 10^{14}$ G for SGR 1900 and $2\times
10^{15}$ G for SGR 1806.  We henceforth take the fiducial values of
$M=1.4M_\odot$ and $R=10$ in each magnetar.  Eigenfrequencies for all $n$
and $l$ for the range of frequencies shown are plotted.  For
$b_t^2=0$, there exists an Alfv\'en continuum that begins at $\simeq
7\,B^d_{15}$ Hz, with a zero-frequency solution corresponding to
rigid-body rotation. As $b_t^2$ is increased, the continuum splits
into discrete normal modes.  As $b_t^2$ is further increased and the
modes spread out further, the higher-frequency modes move off the
diagram.  For large $b_t^2$ the eigenfrequencies become large due to
the large tangled field.

For fixed $M$ and $R$, the model has only $b_t^2$ as a free
parameter. We tune $b_t^2$ to match the 28 Hz QPO observed in SGR 1900,
arriving at $b_t^2=13.3$. Similarly, we tune $b_t^2$ to match the 18 Hz
QPO observed in SGR 1806, arriving at $b_t^2=0.17$. Once $b_t^2$ is
fixed in this way, we have quantitative predictions for the rest of the
mode spectrum. Part of the mode spectrum is shown in Figure
\ref{dots}; the observed QPO frequencies are shown as horizontal 
red dashed lines, and the predicted eigenfrequencies as solid blue
squares. 

The predicted spectrum is given in more detail in Tables
\ref{nu_SGR1900_valuesa} and \ref{nu_SGR1806_values}. 
In Table \ref{nu_SGR1900_valuesa}, we give the spectrum of
eigenfrequencies predicted for SGR 1900. Fundamentals in boldface are
within 3 Hz of an observed QPO, and denote plausible mode
identifications. The predicted spectrum is about three times denser
than the spectrum of observed QPOs. Overtones begin at higher
frequencies, and cannot account for the observed QPO spectrum below 63
Hz.  As we show in \S 3 of the accompanying Letter, overtones are more
energetically costly than the fundamentals. We expect overtones to be
more difficult to observe.

In Table \ref{nu_SGR1806_values}, we give some of the eigenfrequencies
predicted for SGR 1806. The spectrum is very dense, with a mode
spacing of about 2 Hz if we adopt the inferred dipole field of
$2\times 10^{15}$ G. If the dipole field has been overestimated, not
implausible, then the predicted spectrum is much less dense, and more
similar to SGR 1900. For example, taking $B_d$ equal to 0.62 of the
reported value, a value of $b_t^2=1.0$ matches the 18 Hz QPO, and the
predicted spectrum is that given in Table \ref{nu_SGR1806_values1};
the spectrum is less dense, with a spacing of about 7 Hz in the
fundamentals.

\begin{figure*}
\centering
\includegraphics[width=.4\linewidth]{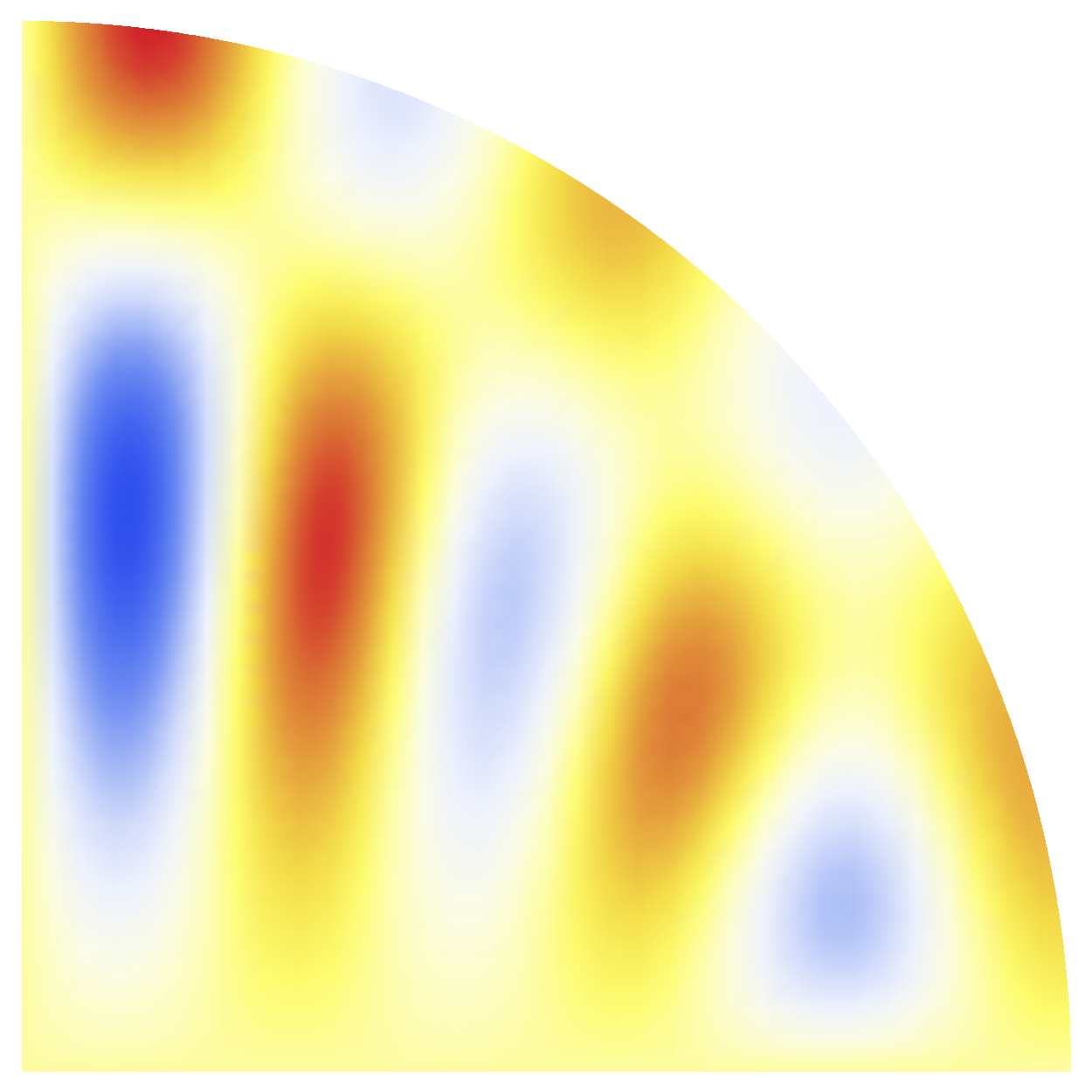} 
\includegraphics[width=.4\linewidth]{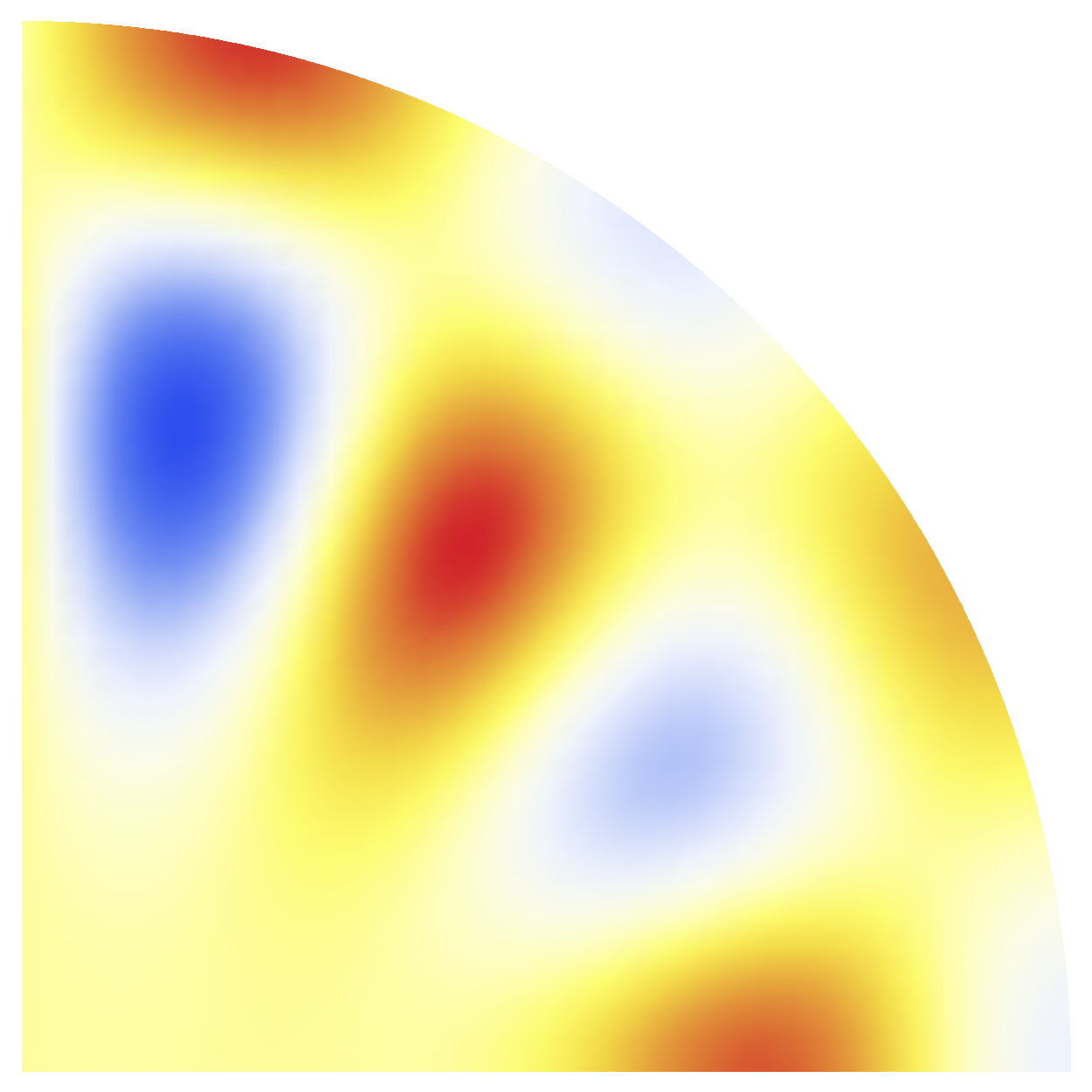} \\
\caption{Example eigenmodes for SGR 1900 (left) and 1806 (right), see text
for details. Light shading denotes motion out of the page for a given
phase, and dark shading denotes motion into the page. The left panel
corresponds to $b_t^2=13.3$ and exhibits the spherical structure of
the isotropic limit, while the right corresponds to $b_t^2=0.17$ and
exhibits the cylindrical structure of the continuum.}
\label{modes}
\end{figure*}

Examples of eigenmodes are plotted in Figure \ref{modes}.  The left
panel shows the 82 Hz mode of SGR 1900, with the attribution $l=6$,
$n=0$, while the right panel shows the 29 Hz mode of SGR 1806, with the
attribution $l=5$, $n=0$. The left panel corresponds to $b_t^2=13.3$, which is
close to the isotropic limit, and shows a large degree of spherical
structure symmetry.  In the plot on the right, $b_t^2=0.17$, and the
mode shows mostly cylindrical structure, a characteristic feature of
the continuum.

All examples shown here are for $R=10$ km and $M=1.4M_\odot$. The
eigenfrequencies scale with the stellar compactness
approximately as $(M/R)^{-1/2}$ for fixed $b_t^2$; see
equation (\ref{nu_isotropic1}). If the compactness is increased,
$b_t^2$ must also be increased to give the same spectrum.

\section{Effects of density stratification and crust rigidity}

\label{stratification}

We now show that the inclusion of density stratification and crust
rigidity give small corrections to the eigenfrequencies obtained with
the constant-density model for $b_t^2>>1$. For simplicity, we ignore the
dipole field, and study the isotropic problem corresponding to a
strong magnetic tangle. In these comparisons, we take the dipole field
to have a strength of $B_d=7\times 10^{14}$ G, as measured for SGR
1900, and $b_t^2=13.3$ inferred above for this object. 

We construct a relativistic star using the analytic representation of
the Brussels-Montreal equation of state derived by
\citet{potekhin_etal13}. For a 1.4 $M_\odot$ neutron star, we obtain a
radius of 13.1 km. The shear modulus is from
\citet{strohmayer_etal91a}, and we fix $x_p=0.1$. 
For this spherical problem, the variable
separation proceeds as in \S \ref{isotropic_model}. We solve
numerically for the radial eigenfunctions and eigenfrequencies.  As a
baseline for comparison, we use equation (\ref{nu_isotropic1}) to obtain the
eigenfrequencies for a constant-density star of radius 13.1 km.

We assess the effects of density stratification by calculating
the change in eigenfrequencies of a stratified star, without a rigid
crust, relative to each eigenfrequency obtained from the
constant-density model of radius $R=13.1$ km and of mass 1.4 $M_\odot$; see
Table \ref{strat_effects}. The fundamental modes all shift up by about
20\%. The overtones go down by about 15\% or less, an effect that
decreases with increasing $l$. 

We assess the effects of crust rigidity by calculating the shift in
eigenfrequencies of a stellar model with a crust relative to the same
model with the material shear modulus set to zero. Crust rigidity
increases the eigenfrequencies by about 4\% or less for the
fundamentals, and about 1\% or less for overtones. These results are in
quantitative agreement with the simple two-zone calculation of these
effects presented in \S 4 of the accompanying Letter. 

In summary, density stratification increases the frequencies of the
fundamental modes by about 20\% relative to the constant-density
model, a shift that can be almost completely subsumed in the
constant-density model with different choices of radius or
mass. Overtones are shifted down by 15\% or less, an effect that
decreases with $l$. The effects of crust rigidity are small enough -
less than 4\% - that the crust can be ignored.

These conclusions hold for $b_t^2>>1$, as we have inferred for SGR
1900. For SGR 1806, with an inferred value of $b_t^2$ of $0.17$, the
effects of the crust could be important, while we expect the effects
of density stratification to be about the same. 

\appendix

\section{Variable Separation}

\label{sov}

Equations (\ref{eomt}) and (\ref{bct}) can be separated and solved by
transforming to an oblate spheroidal coordinate system $(u,v)$ defined
by
\begin{equation}
s=R
\sqrt{\frac{\left(1+b_t^2 u^2\right)\left(1-v^2\right)}
{1+b_t^2}}
\end{equation}
\begin{equation}
z=R u v
\end{equation}
Curves of constant $u$ are ellipses, and curves of constant $v$ are
hyperbolae. For $u=1$, the coordinate gives a sphere of radius
$R$. In the limit $b_t^2\rightarrow \infty$, spherical 
coordinates are recovered with $u=r/R$ and
$v=\cos\theta$. 

In these coordinates, equation (\ref{eomt}) becomes
\begin{eqnarray}
\left(1+b_t^2 u^2\right)\frac{\partial^2 u_\phi}{\partial u^2}+ && 2 b_t^2 u \frac{\partial u_\phi}{\partial u} +\frac{b_t^2 u_\phi}{1+b_t^2 u^2}  
+b_t^2\left(1-v^2\right)\frac{\partial^2 u_\phi}{\partial
v^2}\nonumber \\
&& -2 b_t^2 v \frac{\partial u_\phi}{\partial v} -\frac{b_t^2u_\phi}{1-v^2}  \nonumber 
+\bar{\omega}^2\left(b_t^2 u^2+ v^2\right)u_\phi=0
\label{eomtuv}
\end{eqnarray}
where
\begin{equation}
\bar\omega\equiv \frac{R \omega}{c_t}\sqrt{\frac{b_t^2}{1+b_t^2}}, 
\end{equation}
and $c_t^2\equiv\ev{B_t^2}/(4\pi\rho_d)$.

The boundary condition equation (\ref{bct}) at $u=1$ becomes
\begin{equation}
(1+b_t^2)\frac{\partial u_\phi}{\partial u}-b_t^2 u_\phi=0,
\label{bcuv}
\end{equation}
while at $z=0$ we require
\begin{equation}
  u_\phi=0\,, \label{bczodd}
\end{equation}
for odd parity modes, and 
\begin{equation}
  \left(1+b_t^2 u^2\right) v \frac{\partial u_\phi}{\partial u}+b_t^2\left(1-v^2\right) u \frac{\partial u_\phi}{\partial v}=0\,, \label{bczeven}
\end{equation}
for even parity modes.

We seek a separable solution of the form
\begin{equation}
u_\phi(u,v)=U(u)V(v).
\end{equation}
Equation (\ref{eomtuv}) becomes
\begin{equation}
   \left(1+b_t^2 u^2\right)\frac{d^2 U}{d u^2}+2 b_t^2 u \frac{d U}{d u} +\frac{b_t^2 U}{1+b_t^2 u^2} -\kappa b_t^2 U +\bar{\omega}^2 b_t^2 u^2 U=0
\label{sepu}
\end{equation}
\begin{equation}
b_t^2 \left(1-v^2\right)\frac{d^2 V}{d v^2}-2 b_t^2 v \frac{d V}{d v} -\frac{b_t^2 V}{1-v^2} +\kappa b_t^2 V + \bar{\omega}^2 v^2 V=0,
\label{sepv}
\end{equation}
where $\kappa$ is the separation constant. The boundary condition
eq. (\ref{bcuv}) becomes
\begin{equation}
(1+b_t^2)\frac{dU}{du}-b_t^2U=0,
\label{bcu}
\end{equation}
at $u=1$. 
The boundary conditions (\ref{bczodd}) and (\ref{bczeven}) are satisfied if we impose
\begin{equation}
  U(0)=0 \hspace{5mm} \mbox{and} \hspace{5mm} V(0)=0\,,
\end{equation}
for odd parity modes or
\begin{equation}
  U'(0)=0 \hspace{5mm} \mbox{and} \hspace{5mm} V'(0)=0\,,
\end{equation}
for even parity modes.

In the limit $b_t^2\rightarrow
\infty$, equation (\ref{sepu}) reduces to the spherical Bessel equation
with solution $U(u)=j(\bar{\omega} u)$, while (\ref{sepv}) reduces to
an associated Legendre equation 
with solution
$V(v)=\sqrt{1-v^2}\partial P^0_{l}/\partial v=-P^1_{l}(v)$ and
$\kappa=l(l+1)$, which recovers the solutions of the isotropic model of
\S \ref{isotropic_model}.

We solve equations (\ref{sepu}), (\ref{sepv}), (\ref{bcu}) numerically for
the normal mode frequencies.
For arbitrary  $b_t^2$ the modes are identified by the number of
nodes; for the $n^{th}$ mode the function $U(u)$ has $n$ nodes on the
domain $0\leq u \leq 1$, while for the $l^{th}$ mode the function
$V(v)$ has $l-1$ nodes on $-1\leq v \leq 1$.


\begin{thebibliography}{62}
\expandafter\ifx\csname natexlab\endcsname\relax\def\natexlab#1{#1}\fi

\bibitem[{Barat {et~al.}(1979)Barat, Chambon, Hurley, Niel, Vedrenne, Estulin,
  Kurt, \& Zenchenko}]{barat_etal79}
Barat, C., Chambon, G., Hurley, K., Niel, M., Vedrenne, G., Estulin, I., Kurt,
  V., \& Zenchenko, V. 1979, Astronomy and Astrophysics, 79, L24

\bibitem[{Braithwaite(2008)}]{braithwaite08}
Braithwaite, J. 2008, Monthly Notices of the Royal Astronomical Society, 386,
  1947

\bibitem[{Braithwaite(2009)}]{braithwaite09}
---. 2009, Mon. Not. Roy. Astron. Soc., 397, 763

\bibitem[{Braithwaite \& Nordlund(2006)}]{bn06}
Braithwaite, J. \& Nordlund, {\AA}. 2006, Astronomy \& Astrophysics, 450, 1077

\bibitem[{Braithwaite \& Spruit(2006)}]{bs06}
Braithwaite, J. \& Spruit, H. 2006, Astronomy \& Astrophysics, 450, 1097

\bibitem[{Brandenburg \& Subramanian(2005)}]{brandenburg_etal05}
Brandenburg, A. \& Subramanian, K. 2005, Physics Reports, 417, 1

\bibitem[{Cerd\'a-Dur\'an {et~al.}(2009)Cerd\'a-Dur\'an, Stergioulas, \&
  Font}]{csf09}
Cerd\'a-Dur\'an, P., Stergioulas, N., \& Font, J.~A. 2009, Mon. Not. Roy.
  Astron. Soc., 397, 1607

\bibitem[{Cerd{\'a}-Dur{\'a}n {et~al.}(2009)Cerd{\'a}-Dur{\'a}n, Stergioulas,
  \& Font}]{cerda_etal09}
Cerd{\'a}-Dur{\'a}n, P., Stergioulas, N., \& Font, J.~A. 2009, Monthly Notices
  of the Royal Astronomical Society, 397, 1607

\bibitem[{Chamel(2005)}]{chamel05}
Chamel, N. 2005, Nucl. Phys. A, 747, 109

\bibitem[{Chamel(2012)}]{chamel12}
---. 2012, Phys. Rev. C, 85, 035801

\bibitem[{Chamel \& Haensel(2006)}]{ch06}
Chamel, N. \& Haensel, P. 2006, Phys. Rev. C, 73, 045802

\bibitem[{Cline {et~al.}(1980)Cline, Desai, Pizzichini, Teegarden, Evans,
  Klebesadel, Laros, Hurley, Niel, \& Vedrenne}]{cline_etal80}
Cline, T., Desai, U., Pizzichini, G., Teegarden, B., Evans, W., Klebesadel, R.,
  Laros, J., Hurley, K., Niel, M., \& Vedrenne, G. 1980, The Astrophysical
  Journal, 237, L1

\bibitem[{Colaiuda {et~al.}(2009)Colaiuda, Beyer, \& Kokkotas}]{cbk09}
Colaiuda, A., Beyer, H., \& Kokkotas, K. 2009, Monthly Notices of the Royal
  Astronomical Society, 396, 1441

\bibitem[{Colaiuda \& Kokkotas(2011)}]{ck11}
Colaiuda, A. \& Kokkotas, K.~D. 2011, Mon. Not. Roy. Astron. Soc., 414, 3014

\bibitem[{D'Angelo \& Watts(2012)}]{dw12}
D'Angelo, C. \& Watts, A. 2012, The Astrophysical Journal Letters, 751, L41

\bibitem[{Douchin \& Haensel(2001)}]{dh01}
Douchin, F. \& Haensel, P. 2001, Astron. Astrophys., 380, 151

\bibitem[{{El-Mezeini} \& {Ibrahim}(2010)}]{elib10}
{El-Mezeini}, A.~M. \& {Ibrahim}, A.~I. 2010, Astrophys. J. Lett., 721, L121

\bibitem[{{Esposito} {et~al.}(2010){Esposito}, {Israel}, {Turolla}, {Tiengo},
  {G{\"o}tz}, {de Luca}, {Mignani}, {Zane}, {Rea}, {Testa}, {Caraveo}, {Chaty},
  {Mattana}, {Mereghetti}, {Pellizzoni}, \& {Romano}}]{esposito_etal10}
{Esposito}, P., {Israel}, G.~L., {Turolla}, R., {Tiengo}, A., {G{\"o}tz}, D.,
  {de Luca}, A., {Mignani}, R.~P., {Zane}, S., {Rea}, N., {Testa}, V.,
  {Caraveo}, P.~A., {Chaty}, S., {Mattana}, F., {Mereghetti}, S., {Pellizzoni},
  A., \& {Romano}, P. 2010, \mnras, 405, 1787

\bibitem[{{Fenimore} {et~al.}(1996){Fenimore}, {Klebesadel}, \&
  {Laros}}]{fenimore_etal96}
{Fenimore}, E.~E., {Klebesadel}, R.~W., \& {Laros}, J.~G. 1996, Astrophys. J.,
  460, 964

\bibitem[{Feroci {et~al.}(1999)Feroci, Frontera, Costa, Amati, Tavani,
  Rapisarda, \& Orlandini}]{feroci_etal99}
Feroci, M., Frontera, F., Costa, E., Amati, L., Tavani, M., Rapisarda, M., \&
  Orlandini, M. 1999, The Astrophysical Journal Letters, 515, L9

\bibitem[{Flowers \& Ruderman(1977)}]{fr77}
Flowers, E. \& Ruderman, M. 1977, The Astrophysical Journal, 215, 302

\bibitem[{Gabler {et~al.}(2011)Gabler, Cerd\'a-Dur\'an, Font, M\"uller, \&
  Stergioulas}]{gabler_etal11}
Gabler, M., Cerd\'a-Dur\'an, P., Font, J.~A., M\"uller, E., \& Stergioulas, N.
  2011, Mon. Not. Roy. Astron. Soc., 410, L37

\bibitem[{Gabler {et~al.}(2013{\natexlab{a}})Gabler, Cerd{\'a}-Dur{\'a}n, Font,
  M{\"u}ller, \& Stergioulas}]{gabler_etal13}
Gabler, M., Cerd{\'a}-Dur{\'a}n, P., Font, J.~A., M{\"u}ller, E., \&
  Stergioulas, N. 2013{\natexlab{a}}, Monthly Notices of the Royal Astronomical
  Society, 430, 1811

\bibitem[{Gabler {et~al.}(2012)Gabler, Cerd{\'a}-Dur{\'a}n, Stergioulas, Font,
  \& M{\"u}ller}]{gabler_etal12}
Gabler, M., Cerd{\'a}-Dur{\'a}n, P., Stergioulas, N., Font, J.~A., \&
  M{\"u}ller, E. 2012, Monthly Notices of the Royal Astronomical Society, 421,
  2054

\bibitem[{Gabler {et~al.}(2013{\natexlab{b}})Gabler, Cerd{\'a}-Dur{\'a}n,
  Stergioulas, Font, \& M{\"u}ller}]{gabler_etal_sf13}
---. 2013{\natexlab{b}}, Physical review letters, 111, 211102

\bibitem[{Gabler {et~al.}(2014)Gabler, Cerd{\'a}-Dur{\'a}n, Stergioulas, Font,
  \& M{\"u}ller}]{gabler_etal14}
---. 2014, Monthly Notices of the Royal Astronomical Society, 443, 1416

\bibitem[{Glampedakis {et~al.}(2006)Glampedakis, Samuelsson, \&
  Andersson}]{gsa06}
Glampedakis, K., Samuelsson, L., \& Andersson, N. 2006, Mon. Not. Roy. Astron.
  Soc., 371, L74

\bibitem[{Goedbleod \& Poedts(2004)}]{gp04}
Goedbleod, J. P.~H. \& Poedts, S. 2004, Principles of Magnetohydrodynamics
  (Cambridge Univ. Press, Cambridge)

\bibitem[{{Haensel} \& {Potekhin}(2004)}]{hp04}
{Haensel}, P. \& {Potekhin}, A.~Y. 2004, Astron. Astrophys., 428, 191

\bibitem[{Hambaryan {et~al.}(2011)Hambaryan, Neuh\"auser, \&
  Kokkotas}]{hambaryan_etal11}
Hambaryan, V., Neuh\"auser, R., \& Kokkotas, K.~D. 2011, Astron. Astrophys.,
  528, 45

\bibitem[{Huppenkothen {et~al.}(2014{\natexlab{a}})Huppenkothen, D'Angelo,
  Watts, Heil, van~der Klis, van~der Horst, Kouveliotou, Baring,
  G{\"o}{\u{g}}{\"u}{\c{s}}, Granot, {et~al.}}]{huppenkothen_etal14b}
Huppenkothen, D., D'Angelo, C., Watts, A.~L., Heil, L., van~der Klis, M.,
  van~der Horst, A.~J., Kouveliotou, C., Baring, M.~G.,
  G{\"o}{\u{g}}{\"u}{\c{s}}, E., Granot, J., {et~al.} 2014{\natexlab{a}}, The
  Astrophysical Journal, 787, 128

\bibitem[{Huppenkothen {et~al.}(2014{\natexlab{b}})Huppenkothen, Heil, Watts,
  \& G{\"o}{\u{g}}{\"u}{\c{s}}}]{huppenkothen_etal14a}
Huppenkothen, D., Heil, L., Watts, A., \& G{\"o}{\u{g}}{\"u}{\c{s}}, E.
  2014{\natexlab{b}}, The Astrophysical Journal, 795, 114

\bibitem[{Huppenkothen {et~al.}(2013)Huppenkothen, Watts, Uttley, van~der
  Horst, van~der Klis, Kouveliotou, G{\"o}{\u{g}}{\"u}{\c{s}}, Granot, Vaughan,
  \& Finger}]{huppenkothen_etal13}
Huppenkothen, D., Watts, A.~L., Uttley, P., van~der Horst, A.~J., van~der Klis,
  M., Kouveliotou, C., G{\"o}{\u{g}}{\"u}{\c{s}}, E., Granot, J., Vaughan, S.,
  \& Finger, M.~H. 2013, The Astrophysical Journal, 768, 87

\bibitem[{Hurley {et~al.}(1999)Hurley, Cline, Mazets, Barthelmy, Butterworth,
  Marshall, Palmer, Aptekar, Golenetskii, Il'Inskii, {et~al.}}]{hurley_etal99}
Hurley, K., Cline, T., Mazets, E., Barthelmy, S., Butterworth, P., Marshall,
  F., Palmer, D., Aptekar, R., Golenetskii, S., Il'Inskii, V., {et~al.} 1999,
  Nature, 397, 41

\bibitem[{Israel {et~al.}(2005)Israel, Belloni, Stella, Rephaeli, Gruber,
  Casella, Dall'Osso, Rea, Persic, \& Rothschild}]{israel_etal05}
Israel, G., Belloni, T., Stella, L., Rephaeli, Y., Gruber, D., Casella, P.,
  Dall'Osso, S., Rea, N., Persic, M., \& Rothschild, R. 2005, The Astrophysical
  Journal Letters, 628, L53

\bibitem[{Levin(2006)}]{levin06}
Levin, Y. 2006, Mon. Not. Roy. Astron. Soc., 368, L35

\bibitem[{Levin(2007)}]{levin07}
---. 2007, Mon. Not. Roy. Astron. Soc., 377, 159

\bibitem[{{Link} \& {van Eysden}(2015)}]{lv15}
{Link}, B. \& {van Eysden}, C.~A. 2015, arXiv:1503.01410

\bibitem[{{Link} \& {van Eysden}(2016)}]{lv16s}
---. 2016, ApJ, {\sl in press}, (Supplementary materials)

\bibitem[{Mazets {et~al.}(1979)Mazets, Golenetskii, Il'Inskii, Aptekar, \&
  Guryan}]{mazets_etal79}
Mazets, E., Golenetskii, S., Il'Inskii, V., Aptekar, R., \& Guryan, Y.~A. 1979,
  Nature, 282, 587

\bibitem[{Mereghetti {et~al.}(2006)Mereghetti, Esposito, Tiengo, Zane, Turolla,
  Stella, Israel, G{\"o}tz, \& Feroci}]{mereghetti_etal06}
Mereghetti, S., Esposito, P., Tiengo, A., Zane, S., Turolla, R., Stella, L.,
  Israel, G., G{\"o}tz, D., \& Feroci, M. 2006, The Astrophysical Journal, 653,
  1423

\bibitem[{Nakagawa {et~al.}(2008)Nakagawa, Mihara, Yoshida, Yamaoka, Sugita,
  Murakami, Yonetoku, Suzuki, Nakajima, Tashiro, {et~al.}}]{nakagawa_etal08}
Nakagawa, Y.~E., Mihara, T., Yoshida, A., Yamaoka, K., Sugita, S., Murakami,
  T., Yonetoku, D., Suzuki, M., Nakajima, M., Tashiro, M., {et~al.} 2008, PASJ,
  61, S387

\bibitem[{Passamonti \& Lander(2013)}]{pl13}
Passamonti, A. \& Lander, S. 2013, Monthly Notices of the Royal Astronomical
  Society, 429, 767

\bibitem[{{Rea} {et~al.}(2010){Rea}, {Esposito}, {Turolla}, {Israel}, {Zane},
  {Stella}, {Mereghetti}, {Tiengo}, {G{\"o}tz}, {G{\"o}{\u g}{\"u}{\c s}}, \&
  {Kouveliotou}}]{ret10}
{Rea}, N., {Esposito}, P., {Turolla}, R., {Israel}, G.~L., {Zane}, S.,
  {Stella}, L., {Mereghetti}, S., {Tiengo}, A., {G{\"o}tz}, D., {G{\"o}{\u
  g}{\"u}{\c s}}, E., \& {Kouveliotou}, C. 2010, Science, 330, 944

\bibitem[{{Rea} {et~al.}(2012){Rea}, {Israel}, {Esposito}, {Pons},
  {Camero-Arranz}, {Mignani}, {Turolla}, {Zane}, {Burgay}, {Possenti},
  {Campana}, {Enoto}, {Gehrels}, {G{\"o}{\v g}{\"u}{\c s}}, {G{\"o}tz},
  {Kouveliotou}, {Makishima}, {Mereghetti}, {Oates}, {Palmer}, {Perna},
  {Stella}, \& {Tiengo}}]{rea_etal12}
{Rea}, N., {Israel}, G.~L., {Esposito}, P., {Pons}, J.~A., {Camero-Arranz}, A.,
  {Mignani}, R.~P., {Turolla}, R., {Zane}, S., {Burgay}, M., {Possenti}, A.,
  {Campana}, S., {Enoto}, T., {Gehrels}, N., {G{\"o}{\v g}{\"u}{\c s}}, E.,
  {G{\"o}tz}, D., {Kouveliotou}, C., {Makishima}, K., {Mereghetti}, S.,
  {Oates}, S.~R., {Palmer}, D.~M., {Perna}, R., {Stella}, L., \& {Tiengo}, A.
  2012, \apj, 754, 27

\bibitem[{{Rea} {et~al.}(2014){Rea}, {Vigan{\`o}}, {Israel}, {Pons}, \&
  {Torres}}]{rvi14}
{Rea}, N., {Vigan{\`o}}, D., {Israel}, G.~L., {Pons}, J.~A., \& {Torres}, D.~F.
  2014, \apjl, 781, L17

\bibitem[{{Sotani}(2015)}]{sotani15}
{Sotani}, H. 2015, \prd, 92, 104024

\bibitem[{Sotani {et~al.}(2008{\natexlab{a}})Sotani, Colaiuda, \&
  Kokkotas}]{sotani_etal08a}
Sotani, H., Colaiuda, A., \& Kokkotas, K.~D. 2008{\natexlab{a}}, Mon. Not. Roy.
  Astron. Soc., 385, 2162

\bibitem[{Sotani {et~al.}(2008{\natexlab{b}})Sotani, Kokkotas, \&
  Stergioulas}]{sotani_etal08b}
Sotani, H., Kokkotas, K.~D., \& Stergioulas, N. 2008{\natexlab{b}}, Mon. Not.
  Roy. Astron. Soc., 385, L5

\bibitem[{Strohmayer {et~al.}(1991)Strohmayer, van Horn, Ogata, Iyetomi, \&
  Ichimaru}]{strohmayer_etal91}
Strohmayer, T.~E., van Horn, H.~M., Ogata, S., Iyetomi, H., \& Ichimaru, S.
  1991, Astrophys. J., 375, 679

\bibitem[{Strohmayer \& Watts(2005)}]{sw05}
Strohmayer, T.~E. \& Watts, A.~L. 2005, Astrophys. J., 632, L111

\bibitem[{Strohmayer \& Watts(2006)}]{sw06}
---. 2006, The Astrophysical Journal, 653, 593

\bibitem[{Tayler(1973)}]{tayler73}
Tayler, R. 1973, Monthly Notices of the Royal Astronomical Society, 161, 365

\bibitem[{Thompson \& Duncan(1993)}]{td93}
Thompson, C. \& Duncan, R.~C. 1993, Astrophys. J., 408

\bibitem[{Tiengo {et~al.}(2009)Tiengo, Esposito, Mereghetti, Israel, Stella,
  Turolla, Zane, Rea, G{\"o}tz, \& Feroci}]{tiengo_etal09}
Tiengo, A., Esposito, P., Mereghetti, S., Israel, G., Stella, L., Turolla, R.,
  Zane, S., Rea, N., G{\"o}tz, D., \& Feroci, M. 2009, Mon. Not. Roy. Astron.
  Soc., 399, L74

\bibitem[{Timokhin {et~al.}(2008)Timokhin, Eichler, \&
  Lyubarsky}]{timokhin_etal08}
Timokhin, A., Eichler, D., \& Lyubarsky, Y. 2008, The Astrophysical Journal,
  680, 1398

\bibitem[{Turolla {et~al.}(2011)Turolla, Zane, Pons, Esposito, \&
  Rea}]{turolla_etal11}
Turolla, R., Zane, S., Pons, J., Esposito, P., \& Rea, N. 2011, The
  Astrophysical Journal, 740, 105

\bibitem[{{van der Horst} {et~al.}(2010){van der Horst}, {Connaughton},
  {Kouveliotou}, {G{\"o}{\v g}{\"u}{\c s}}, {Kaneko}, {Wachter}, {Briggs},
  {Granot}, {Ramirez-Ruiz}, {Woods}, {Aptekar}, {Barthelmy}, {Cummings},
  {Finger}, {Frederiks}, {Gehrels}, {Gelino}, {Gelino}, {Golenetskii},
  {Hurley}, {Krimm}, {Mazets}, {McEnery}, {Meegan}, {Oleynik}, {Palmer},
  {Pal'shin}, {Pe'er}, {Svinkin}, {Ulanov}, {van der Klis}, {von Kienlin},
  {Watts}, \& {Wilson-Hodge}}]{horst_etal10}
{van der Horst}, A.~J., {Connaughton}, V., {Kouveliotou}, C., {G{\"o}{\v
  g}{\"u}{\c s}}, E., {Kaneko}, Y., {Wachter}, S., {Briggs}, M.~S., {Granot},
  J., {Ramirez-Ruiz}, E., {Woods}, P.~M., {Aptekar}, R.~L., {Barthelmy}, S.~D.,
  {Cummings}, J.~R., {Finger}, M.~H., {Frederiks}, D.~D., {Gehrels}, N.,
  {Gelino}, C.~R., {Gelino}, D.~M., {Golenetskii}, S., {Hurley}, K., {Krimm},
  H.~A., {Mazets}, E.~P., {McEnery}, J.~E., {Meegan}, C.~A., {Oleynik}, P.~P.,
  {Palmer}, D.~M., {Pal'shin}, V.~D., {Pe'er}, A., {Svinkin}, D., {Ulanov},
  M.~V., {van der Klis}, M., {von Kienlin}, A., {Watts}, A.~L., \&
  {Wilson-Hodge}, C.~A. 2010, \apjl, 711, L1

\bibitem[{van Hoven \& Levin(2011)}]{vl11}
van Hoven, M. \& Levin, Y. 2011, Mon. Not. Roy. Astron. Soc., 420, 1036

\bibitem[{van Hoven \& Levin(2012)}]{vl12}
---. 2012, Monthly Notices of the Royal Astronomical Society, 420, 3035

\bibitem[{Watts \& Strohmayer(2006)}]{ws06}
Watts, A.~L. \& Strohmayer, T.~E. 2006, Astrophys. J., 637, L117

\bibitem[{{Wright}(1973)}]{wright73}
{Wright}, G.~A.~E. 1973, Mon. Not. Roy. Astron. Soc., 162, 339

\end{thebibliography}

\begin{thebibliography}{4}
\expandafter\ifx\csname natexlab\endcsname\relax\def\natexlab#1{#1}\fi

\bibitem[{Chamel \& Haensel(2006)}]{ch06a}
Chamel, N. \& Haensel, P. 2006, Phys. Rev. C, 73, 045802

\bibitem[{Levin(2007)}]{levin07a}
Levin, Y. 2007, Mon. Not. Roy. Astron. Soc., 377, 159

\bibitem[{Potekhin {et~al.}(2013)Potekhin, Fantina, Chamel, Pearson, \&
  Goriely}]{potekhin_etal13}
Potekhin, A., Fantina, A., Chamel, N., Pearson, J., \& Goriely, S. 2013,
  Astronomy \& astrophysics, 560, A48

\bibitem[{Strohmayer {et~al.}(1991)Strohmayer, van Horn, Ogata, Iyetomi, \&
  Ichimaru}]{strohmayer_etal91a}
Strohmayer, T.~E., van Horn, H.~M., Ogata, S., Iyetomi, H., \& Ichimaru, S.
  1991, Astrophys. J., 375, 679

\end{thebibliography}

\begin{table}
\begin{tabular}{l|llllll}
\hline
$l$ & $n=0$ & $n=1$ & $n=2$ & $n=3$ & \\
\hline
1 & 0 & 63 & 99 & 134 \\
2 & {\bf 28} & 79 & 116 & 152 \\
3 & 43 & 93 & 130 &    \\
4 & {\bf 56} & 107 & 146 &    \\
5 & 69 &  120 & 160 &    \\
6 & {\bf 82} & 134 & &    \\
7 & 94 & 147 & &    \\
8 & 106 & &  &    \\
9 & 118 &  &  &    \\
10 & 130 &  &  &    \\
11 & 142 &  &  &    \\
12 & {\bf 154} &  &  &    
\end{tabular}
\caption{Predicted eigenfrequencies in Hz for SGR 1900 with $B_d=7\times
10^{14}$ G and $b_t^2=13.3$. Fundamentals in boldface are within 3 Hz
of an observed QPO, and denote plausible mode identifications. 
We do not list frequencies above 160 Hz.}
\label{nu_SGR1900_valuesa}
\end{table}

\begin{table}
\begin{tabular}{l|lllll}
\hline
$l$ & $n=0$ & $n=1$ & $n=2$ & $n=3$ & $n=4$ \\
\hline
1 & 0   & 35 & 65 & 93   & 122   \\
2 & 20 & 50 & 79 & 108  & 135  \\ 
3 & 18 & 44 & 72 & 101  & 130   \\
4 & 28 & 58 & 87 & 115  & 143   \\
5 & 29 & 52 & 80 & 109 & 137  \\
6 & 38 & 66 & 95 & 123 & 151  \\
7 & 40 & 62 & 89 & 117 & 145  \\
8 & 48 & 75 & 102 & 131  & 159 \\
9 & 51 & 71 & 97 & 125  & 153   \\
10 & 58 & 84 & 111 & 139  \\
\end{tabular}
\caption{Some of the eigenfrequencies in Hz predicted for SGR 1806
assuming, $B_d=2\times 10^{15}$ G. The 18 Hz QPO requires 
$b_t^2=0.17$, giving a very dense spectrum. }
\label{nu_SGR1806_values}
\end{table}

\begin{table}
\begin{tabular}{l|lllll}
\hline
$l$ & $n=0$ & $n=1$ & $n=2$ & $n=3$ & $n=4$ \\
\hline
1 & 0   & 34 & 58 & 81   & 104   \\
2 & {\bf 18} & 46 & 70 & 93  & 116  \\ 
3 & {\bf 25} & 49 & 72 & 94  & 117   \\
4 & {\bf 32} & 59 & 82 & 106  & 129   \\
5 & 39 & 65 & 85 & 107 & 130  \\
6 & 46 & 74 & 96 & 119 & 142  \\
7 & 53 & 80 & 100 & 121 & 143  \\
8 & {\bf 60} & 88 & 110 & 132  & 155 \\
9 & 66 & 95 & 115 & 135  & 157   \\
10 & 73 & 102 & 124 & 146  \\
\end{tabular}
\caption{Some of the eigenfrequencies in Hz predicted for SGR 1806
assuming, $B_d=1.24\times 10^{15}$ G. The 18 Hz QPO requires 
$b_t^2=1.0$. Fundamentals in boldface are within 3 Hz
of an observed QPO, and denote plausible mode identifications. 
We do not list frequencies above 160 Hz.}
\label{nu_SGR1806_values1}
\end{table}

\begin{table}
\begin{tabular}{l|llll}
\hline
$l$ & $n=0$ & $n=1$ & $n=2$ & $n=3$  \\
\hline
1 &  -         & -15\%& -15\%& -15\% \\
2 & +21\%& -12\%& -14\%& -15\%\\
3 & +21\%& -9.8\%& -13\%&          \\
4 & +21\%& -7.8\%& -12\%&          \\
5 & +21\%& -6.1\%& -10\%&          \\
6 & +20\%& -4.8\%& -10\%&          \\
7 & +20\%& -3.7\%&           &           \\
8 &            & -2.7\%&           & 
\end{tabular}
\caption{Percentage frequency change of modes in a stratified star
with no crust shear modulus compared with a constant density star with
the same mass and 
radius.  
We use a 1.4 $M_\odot$ star with radius 13.1 km, taking $x_p=0.1$,
$B_d=7\times 10^{14}$ G, and $b_t^2=13.3$.}
\label{strat_effects}
\end{table}

\begin{table}
\begin{tabular}{l|lllllll}
\hline
$l$ & $n=0$ & $n=1$ & $n=2$ & $n=3$  \\
\hline
1 &  -          & 0.0     \%& 0.0\%& 0.0\% \\
2 & +4.3\%& 0.61\%& 0.36\%& 0.17\%\\
3 & +4.1\%& 0.93\%& 0.57\%&          \\
4 & +3.4\%& 1.2\%& 0.76\%&          \\
5 & +2.8\%& 1.5\%& 0.86\%&          \\
6 & +2.2\%& 1.6\%& 1.2\%&          \\
7 & +1.8\%&           &           &           
\end{tabular}
\caption{Percentage frequency change of modes in a stratified star
with a crust compared with a stratified star with no crust shear
modulus, for the parameters of Table \ref{strat_effects}.  }
\label{crust_values}
\end{table}

\end{document}